\documentclass[twocolumn,twocolappendix]{aastex631}
\usepackage{graphicx}
\usepackage{dcolumn}
\usepackage{bm}
\usepackage{amsfonts}
\usepackage{color}
\usepackage{newtxtext,amsmath}
\usepackage{appendix}
\usepackage{subfigure}
\usepackage{soul}

\usepackage{tabularx}

\begin{document}

\title{Constraining the Onset Density for the QCD Phase Transition with the Neutrino Signal from  Core-collapse Supernovae}

\correspondingauthor{Noshad Khosravi Largani}

\author[0000-0003-1551-0508]{Noshad Khosravi Largani$^*$}
\affiliation{Institute of Theoretical Physics, University of Wroclaw, Plac Maksa Borna 9, 50-204 Wroclaw, Poland; 
{ \color{blue} noshad.khosravilargani@uwr.edu.pl, tobias.fischer@uwr.edu.pl }}

\author[0000-0003-2479-344X]{Tobias Fischer$^\dagger$}
\affiliation{Institute of Theoretical Physics, University of Wroclaw, Plac Maksa Borna 9, 50-204 Wroclaw, Poland;
{ \color{blue} noshad.khosravilargani@uwr.edu.pl, tobias.fischer@uwr.edu.pl }}

\author[0000-0001-9793-240X]{Niels-Uwe F. Bastian}
\affiliation{Institute of Theoretical Physics, University of Wroclaw, Plac Maksa Borna 9, 50-204 Wroclaw, Poland;
{ \color{blue} noshad.khosravilargani@uwr.edu.pl, tobias.fischer@uwr.edu.pl }}

\received{{\em 2023 May 11}}
\revised{{\em 2024 January 27}}
\accepted{{\em 2024 January 29}} 
\published{{\em 2024 March 27}}

\begin{abstract}
The occurrence of a first-order hadron-quark matter phase transition at high baryon densities is investigated in astrophysical simulations of core-collapse supernovae, to decipher yet incompletely understood properties of the dense matter equation of state (EOS) using neutrinos from such cosmic events. 
It is found that the emission of a nonstandard second neutrino burst, dominated by electron antineutrinos, is not only a measurable signal for the appearance of deconfined quark matter but also reveals information about the state of matter at extreme conditions encountered at the supernova (SN) interior. 
To this end, a large set of spherically symmetric SN models is investigated, studying the dependence on the EOS and the stellar progenitor. 
General relativistic neutrino-radiation hydrodynamics is employed featuring three-flavor Boltzmann neutrino transport and a microscopic hadron-quark hybrid matter EOS class.
 Therefore, the DD2 relativistic mean-field hadronic model is employed, and several variations of it, and the string-flip model for the description of deconfined quark matter.
The resulting hybrid model
covers a representative range of onset densities for the phase transition and latent heats. 
This facilitates the direct connection between intrinsic signatures of the neutrino signal and properties of the EOS. 
In particular, a set of linear relations has been found empirically. 
These potentially provide a constraint for the onset density of a possible QCD phase transition from the future neutrino observation of the next galactic core-collapse SN, if a millisecond electron anti-neutrino burst is present around or less than 1s.

\smallskip

\small {\em Unified Astronomy Thesaurus concepts:} {\color{blue}Supernova dynamics (1664), Supernova neutrinos (1666), Compact objects (288), High energy astrophysics (739), Hydrodynamics (1963)} \\ \\
\end{abstract}

\section{Introduction}
\label{sec:intro}
Stars more massive than about 9~M$_\odot$ end their life as a core-collapse supernova (CCSN). Thereby, a hydrodynamic shock-wave forms. It first propagates quickly outward but stalls later due to the energy losses from the release of the $\nu_e$ burst, which is associated with the shock propagation across the $\nu_e$ neutrinosphere, and the dissociation of nuclei from the collapsing outer layers of the stellar core. The SN problem is related to the revival of the stalled bounce shock through the transfer of energy from the central proto-neutron star (PNS) into the post-shock layer \citep[c.f.][and references therein]{langanke03,Janka07,mirizzi16,Burrows21}. Several scenarios for the shock revival have been proposed, the magneto-rotational mechanism by \citet{LeBlancWilson70}, the acoustic mechanism by \citet{Burrows06b}, and the currently considered standard neutrino heating mechanism by \citet{Bethe85}. In addition, a fourth mechanism has been proposed by \citet{Sagert09}, due to a first-order phase transition, from normal nuclear (in general hadronic) matter, where quarks and gluons are confined to hadrons, to deconfined quark matter. 

Thereby, the essential aspect is the presence of an instability in the hadron-quark matter coexistence region, with a significantly reduced polytropic index. It arises from the commonly employed two-phase approach, with separate equations of state (EOSs) for hadronic and quark matter, and the subsequent phase transition construction. This causes the PNS to collapse with the formation of a second shock wave. It forms when the pure quark matter phase is reached where the EOS stiffens, i.e. the polytropic index increases. The initial propagation of the second shock to increasingly larger radii, and taking over the standing bounce shock, initiates the SN explosion. The previously employed bag model EOS in \citet{Sagert09}, and later in \citet{Fischer11}, is incompatible with observations of massive pulsars of about 2~M$_\odot$ \citep[][]{Antoniadis13,Fonseca:2021}. This caveat has been overcome with the development of a microscopic quark matter EOS in \citet{Kaltenborn17}, with the implementation of repulsive interactions \citep[see also][]{Benic:2015,Klaehn:2015,Klaehn:2017} and a mechanism that mimics confinement through divergent quark masses. The extension of this EOS to finite temperatures and arbitrary isospin asymmetry gave rise to SN explosions of progenitor stars in the zero-age main-sequence (ZAMS) mass range of $30$--$75$~M$_\odot$ \citep[see][]{Fischer18,Fischer20,Fischer:2021,Kuroda2022}. 

One observable feature that all these SN simulations have in common is the release of a second millisecond neutrino burst \citep[see also][]{Zha20,JakobusMueller2022,Lin2024PhRvD.109b3005L}, associated with the propagation of the second shock across the neutrinospheres, with a certain delay after the $\nu_e$ bounce burst. Unlike the latter, the second neutrino burst is dominated by electron antineutrinos, which have the largest detection prospects within the current generation of water Cherenkov detectors through the inverse beta decay. Furthermore, such a phase transition has a distinct gravitational wave signal \citep[see][]{Zha20,Kuroda2022,Zha22,JakobusMueller2023}.  

The present paper extends the previous analyses of SN driven by a first-order hadron-quark phase transition, establishing a novel connection between properties of the observable second neutrino burst and gross thermodynamics properties of the hadron-quark matter hybrid EOS. The phase transition is investigated systematically, employing the newly developed relativistic density functional (RDF) model of \citet{Bastian:2021}. This approach enables us to constrain the onset density of the QCD phase transition using the SN neutrino signal, if a second neutrino burst is present around or less than 1s after the onset of neutrino detection. It is complementary to the findings in \citet{Blacker20}, where a (lower) bound for the possible onset density of quark matter was deduced from the gravitational wave signals of binary neutron star mergers featuring a first-order phase transition \citep[see also][]{Bauswein19,Most19,Bauswein20}.

This manuscript is organized as follows. In sec.~\ref{sec:model} our CCSN model will be reviewed briefly together with the hadron-quark hybrid EOS, followed by the progenitor discussion in sec.~\ref{sec:prog}. The systematic variations of the impact of QCD phase transition in SN simulations are discussed in sec.~\ref{sec:systematics} and the subsequent neutrino signatures are elaborated in sec.~\ref{sec:neutrinos}. The manuscript closes with a summary in \ref{sec:summary}. 

\begin{table*}
\caption{Neutrino Reactions Considered, Including References.}
\begin{tabularx}{1.\textwidth}{c@{\hspace{1cm}}c@{\hspace{2cm}}l}
\hline
\hline
& Weak Process & Reference \\
\hline
1 & $e^- + p \rightleftarrows n + \nu_e$ & \citet{Guo2020PhRvD102} \\ 
2 & $e^+ + n \rightleftarrows p + \bar\nu_e$ & \citet{Guo2020PhRvD102} \\
3 & $n \rightleftarrows p + e^- + \bar\nu_e$ & \citet{Fischer2020PhRvC101} \\
4 & $\nu_e + (A,Z-1) \rightleftarrows (A,Z) + e^-$ & \citet{Juodagalvis:2010} \\
5 & $\nu + N \rightleftarrows \nu' + N$ & \citet{Bruenn85,Mezzacappa93a,horowitz02} \\
6 & $\nu + (A,Z) \rightleftarrows \nu' + (A,Z)$ & \citet{Bruenn85,Mezzacappa93a} \\
7 & $\nu + e^\pm \rightleftarrows \nu' + e^\pm$ & \citet{Bruenn85,Mezzacappa93c} \\
8 & $e^- + e^+ \rightleftarrows  \nu + \bar{\nu}$ & \citet{Bruenn85} \\
9 & $N + N \rightleftarrows  \nu + \bar{\nu} + N + N $ & \citet{hannestad98,Fischer16} \\
10 & $\nu_e + \bar\nu_e \rightleftarrows  \nu_{\mu/\tau} + \bar\nu_{\mu/\tau}$ & \citet{Buras06a,Fischer09} \\
\hline
\end{tabularx}
\\
\\
{\bf Note. } $\nu=\{\nu_e,\bar{\nu}_e,\nu_{\mu/\tau},\bar{\nu}_{\mu/\tau}\}$ and $N=\{n,p\}$
\label{tab:nu-reactions}
\end{table*}
%

\section{CCSN model with QCD phase transition}
\label{sec:model}
For the present study of the hadron-quark phase transition in CCSN, we employ the general relativistic neutrino radiation hydrodynamics model in spherical symmetry {\tt AGILE-BOLTZTRAN} \citep[][]{Mezzacappa93a,Mezzacappa93b,Mezzacappa93c}. It features an adaptive baryon mass mesh \citep[][]{Liebendorfer01PhRvD63,Fischer09}  
for which we employ here 207 grid points. 
For the neutrino transport, the general relativistic Boltzmann equation for three neutrino flavors is solved assuming massless particles. 
The weak reactions used are listed in Table~\ref{tab:nu-reactions}, together with the corresponding references for the implementation of the rates. 
These include the electronic charged current processes (1) and (2) for which the fully inelastic rates of \cite{Guo2020PhRvD102} are used, the (inverse) neutron decay (3), also based on the inelastic treatment implemented in \citet{Fischer2020PhRvC101}, both include weak magnetism contributions, further nuclear electron captures based on the shell model calculations of \citet{Juodagalvis:2010}, neutrino-nucleus and neutrino-nucleon scattering~(5) and (6) within the elastic approximation of \citet{Bruenn85},
including approximate contributions for nucleon recoil and weak magnetism following \citet{horowitz02},
inelastic neutrino electron and positron scattering~(7) of \citet{Bruenn85} and \citet{Mezzacappa93c} and further a variety of pair processes~(8)--(10), such as the standard electron-positron annihilation \citep[][]{Bruenn85}, nucleon-nucleon bremsstrahlung \citep[][]{hannestad98} with leading order medium modifications based on \citet{Fischer16} as well as electron neutrino pair annihilation \citep[][]{Buras06a} following the implementation of \citet{Fischer09}. 
Further details about {\tt AGILE-BOLTZTRAN}, in particular, about the finite differencing representations, can be found in \citet{Liebendorfer04}. \\

{\tt AGILE-BOLTZTRAN} employs a flexible EOS module, which was implemented in \citet{Hempel12}. For the hadronic part, this includes the routines of the Skyrme EOS of \citet{LSEOS} as well as the entire set of relativistic mean field (RMF) EOS tabulations provided by \citet{Hempel2010NuPhA837}. The latter features a medium-modified low-density and low-temperature nuclear statistical equilibrium (NSE) model for several thousand nuclear species. The transition to homogeneous nuclear matter is modeled via an excluded volume approach. Furthermore, the low-temperature part of the SN domain is taken into account assuming silicon-sulfur composition, precisely when the temperature is below $T\leq 0.45$~MeV. This corresponds to the NSE to non-NSE transition in the nuclear reaction networks of stellar evolution calculations. 
 Hence, it is in accordance with the stellar progenitor composition, as will be discussed further below. Additional contributions from electrons, positrons, photons, and Coulomb correlations are added following \citet{Timmes1999}. 

\subsection{Hadron-quark Hybrid EOS}
\label{sec:eos}
Hadron-quark matter hybrid EOS are employed with first-order phase transition, from the class of microscopic RDF EOS of \citet{Fischer18} and \citet{Bastian:2021} is implemented into {\tt AGILE-BOLTZTRAN}. 
Quark matter is modeled using the RDF approach of \citet{Kaltenborn17}, where confinement has been taken into account by introducing string-like quark-quark interaction in the scalar self-energy \citep[c.f.][ and references therein]{Horowitz1985PhRvD31_stringflip,Ropke86}. 
Furthermore, linear and higher-order repulsive interactions are taken into account following \citet{Benic:2015}, which is based on the quasi-particle approach, known from Nambu--Jona-Lasinio models \citep[][]{NJL:1961,Buballa2005PhR} as well as the class of vector interaction enhanced bag models of \citet{Klaehn:2015}. 
These give rise to additional pressure contributions with increasing density and are hence essential for the stability of compact hybrid stars (neutron stars with quark matter cores) in agreement with the high-precision pulsar observations \citep[c.f.][and references therein]{Antoniadis13,Fonseca:2021}. 
Furthermore, the RDF EOS includes a $\rho$-meson equivalent term that controls, in particular, the isospin asymmetry dependence of the hadron-quark phase transition. 
The values for the different nine RDF parameterizations are listed in Table~I of \citet{Bastian:2021}, including the $\rho$-meson term. The latter are selected to minimize the jump of the electron fraction across the phase transition boundary. 

Different representations of the original hadronic DD2 EOS from \citet{Typel10} have been implemented for the RDF quark matter model's phase transition construction. The flow constrain corrected DD2 EOS \citep[see][]{Danielewicz:2002}, known as DD2F \citep[][]{Alvarez-Castillo16} is used for RDF-1.1 and 1.3 -- 1.7. 
The DD2F EOS experiences a substantial softening of the high-density phase, in excess of about twice nuclear saturation density, compared to DD2. 
The saturation density for the DD2F hadronic EOS is $n_{\rm sat}=0.145$~fm$^{-3}$ or equivalently $\rho_{\rm sat}=2.44\times 10^{14}$~g~cm$^{-3}$.
For RDF-1.2 an excluded volume modification, DD2Fev, is used. The latter results in the deviation from the DD2F EOS at densities in excess of about $\rho\simeq 3\times 10^{14}$~g~cm$^{-3}$. For the RDF-1.8 and RDF-1.9 models, both of which feature the lowest onset densities for the first-order phase transition, the original stiff DD2 EOS was employed.

\begin{table*}[htp]
\caption{Thermodynamic Properties of the RDF Class of Hadron-quark Hybrid EOS, Assuming $\beta$-equilibrium for the Zero-temperature Configurations and Constant Electron Lepton Number of $Y_{\rm L}=0.3$ for the Finite Entropy Cases$^a$.}
\begin{tabularx}{0.99\textwidth}{l@{\hspace{1.cm}}c@{\hspace{0.5cm}}c@{\hspace{0.5cm}}c@{\hspace{1.cm}}c@{\hspace{1.5cm}}c@{\hspace{1.5cm}}l}
\hline
\hline
EOS & Condition & $\rho_{\rm onset}^b$ & $\rho_{\rm final}^c$ & $M_{\rm onset}^d$ & $M_{\rm max}^e$ & $Q^f$ \\
 Hadronic \; Quark & & $[10^{14}\,\,\,{\rm g\,\,\,cm}^{-3}]$ & $[10^{14}\,\,\,{\rm g\,\,\,cm}^{-3}]$ & $[{\rm M}_\odot]$ & $[{\rm M}_\odot]$ &  \\
\hline
DD2F \; RDF-1.1 & $T=0$ & 8.8 & 10.5 & 1.55 & 2.13 &  0.90 \\ 
DD2F \; RDF-1.1 & $s=3~k_{\rm B}$ & 6.1 & 10.0 & 1.64 & 2.12 &  2.82 \\ 
\hline
DD2Fev \; RDF-1.2 & $T=0$ & 7.3 & 8.6 & 1.37 & 2.15 &  0.78 \\ 
DD2Fev \; RDF-1.2 & $s=3~k_{\rm B}$ & 4.9 & 7.8 & 1.45 & 2.17 &  3.95 \\
\hline
DD2F \; RDF-1.3 & $T=0$ & 9.0 & 10.4 & 1.56 & 2.02 &  0.75 \\ 
DD2F \; RDF-1.3 & $s=3~k_{\rm B}$ & 6.0 & 9.5 & 1.63 & 2.03 &  3.71 \\
\hline
DD2F \; RDF-1.4 & $T=0$ & 9.7 & 11.0 & 1.66 & 2.02 &  0.74 \\ 
DD2F\;  RDF-1.4 & $s=3~k_{\rm B}$ & 6.2 & 9.9 & 1.69 & 2.02 &  3.44 \\ 
\hline
DD2F \; RDF-1.5 & $T=0$ & 8.2 & 9.9 & 1.46 & 2.03 &  0.85 \\ 
DD2F \; RDF-1.5 & $s=3~k_{\rm B}$ & 5.5 & 9.5 & 1.57 & 2.04 &  2.38 \\ 
\hline
DD2F \; RDF-1.6 & $T=0$ & 9.0 & 11.0 & 1.58 & 2.00 &  0.92 \\ 
DD2F \; RDF-1.6 & $s=3~k_{\rm B}$ & 6.1 & 10.2 & 1.67 & 2.01 & 3.07 \\
\hline
DD2F \; RDF-1.7 & $T=0$ & 8.9 & 9.8 & 1.61 & 2.11 &  0.49 \\ 
DD2F \; RDF-1.7 & $s=3~k_{\rm B}$ & 5.4 & 8.6 & 1.57 & 2.12 & 2.22 \\ 
\hline
DD2 \; RDF-1.8 & $T=0$ & 4.8 & 8.6 & 0.95 & 2.04 &  1.85 \\ 
DD2 \; RDF-1.8 & $s=3~k_{\rm B}$ & 3.8 & 8.4 & 1.35 & 2.09 & 2.90 \\ 
\hline
DD2 \; RDF-1.9 & $T=0$ & 4.6 & 7.5 & 0.81 & 2.16 &  1.48 \\ 
DD2 \; RDF-1.9 & $s=3~k_{\rm B}$ & 3.3 & 7.4 & 1.25 & 2.21 & 4.35 \\ 
\hline
\end{tabularx}
\\
\\
{\bf Notes.}\\
$^a$~EOS data from \citet{Khosravi:2022} \\
$^b$~Onset density for the phase transition, determined when a finite value of the quark volume fraction is obtained, $\chi_{\rm quark}>0$ \\
$^c$~Density for reaching the pure quark matter phase, where $\chi=1$ \\
$^d$~Onset mass of the hybrid branches \\
$^e$~Maximum mass of the corresponding hybrid EOS \\
$^f$~ Latent heat (see text for definition)
\label{tab:EOS}
\end{table*}

The phase transition construction for this two-phase approach, from the hadronic DD2/DD2F/DD2Fev EOS to the quark matter RDF EOS, is modeled via the assumption of mechanical equilibrium, realized through equal pressures in both phases and chemical phase equilibrium in which the chemical potentials equal in both phases. 
This is known as Maxwell construction. 
However, since under SN conditions, there are two conserved currents, namely, the baryon and the charge numbers, given in terms of the rest-mass density (the following relation holds between rest-mass density and baryon density, $\rho=m_{\rm B}n_{\rm B}$, for which we assume in {\tt AGILE-BOLTZTRAN} $m_{\rm B}=938$~MeV) and the leptonic charge fraction, under the assumption of charge neutrality, $(\rho, Y_e)$. 
Hence, we are dealing with two associated {baryon and hadronic charge} chemical potentials, $(\mu_{\rm B},\mu_{\rm Q})$. 
For practical purposes, the phase transition construction is performed for a constant value of the charge chemical potential, in order to determine pressure equilibrium with respect to the baryon chemical potential. However, since the multipurpose astrophysical hadronic EOSs are provided in terms of densities, they are first mapped from densities $(\rho, Y_e)$ to the chemical potentials $(\mu_{\rm B},\mu_{\rm Q})$, for the calculation of the phase transition construction, and later mapped back from chemical potentials to the densities. 
This introduces truncation errors that must be monitored well. 
Furthermore, the Maxwell conditions result in a jump of all thermodynamic quantities, at the onset of the phase transition from the hadronic phase to the quark matter phase. 
However, in astrophysical simulations, data must be provided also in the density domain in between the two hadronic and quark matter phases. 
Therefore, a quark matter volume fraction, denoted as $\chi_{\rm quark}$, is defined \citep[for details, see][]{Bastian:2021}, based on which the hadron-quark mixed phase is designed. 
 It takes values of $\chi_{\rm quark}\in [0,1]$, indicating nuclear matter ($\chi_{\rm quark}=0$), quark matter ($\chi_{\rm quark}=1$) as well as the hadron-quark mixed phase in between ($0<\chi_{\rm quark}<1$). 
Typical for a first-order phase transition is the sudden change in slope, i.e. the sudden drop in the polytropic index at the transition from the hadronic to the mixed phase. 
This feature is particularly present for the zero-temperature EOS, while it is substantially milder for the finite temperature, more precisely the finite entropy case, explored here for $s=3~k_{\rm B}$. 
Selected properties of the phase transition for all nine RDF EOS are listed in Table~\ref{tab:EOS}, 
these include the onset densities and PNS masses for the phase transition, $\rho_{\rm onset}$ and $M_{\rm onset}$, as well as the densities for reaching the pure quark matter phase $\rho_{\rm final}$ and the maximum masses $M_{\rm max}$. 

In addition, we compute the latent heat, for which we follow closely the definition based on the energy densities of \citet{Lope-Oter2022PhRvC.105e2801L}, Eq.~(6) denoted as $L\vert_\varepsilon$, slightly modified taking into account pressure differences in the hadronic and quark phases, $P_{\rm onset}$ and $P_{\rm final}$, respectively, for the finite entropy configurations, as follows:
\begin{eqnarray}
Q = \frac{P_{\rm final}\varepsilon_{\rm final}-P_{\rm onset}\varepsilon_{\rm onset}}{\varepsilon_{\rm onset}\varepsilon_{\rm final}}~,
\end{eqnarray}
with $\varepsilon_{\rm onset}$ and $\varepsilon_{\rm final}$ defined as the energy density for the onset of the phase transition and the onset of pure quark matter, respectively
The values we find at $T=0$ are in the range of $Q=0.5-1.9$, which is in agreement with the values concluded in \citet{Lope-Oter2022PhRvC.105e2801L}, $Q>0.1$ for neutron star matter.
We determine these values, starting at low densities, where $\chi>0$ for the first time for $P_{\rm onset}$ and $\varepsilon_{\rm onset}$, and after that where $\chi=1$ for $P_{\rm final}$ and $\varepsilon_{\rm final}$, respectively.
The extended density jumps of the hadron-quark mixed phase obtained for finite temperatures, mostly due to lower onset densities of the mixed phase \citep[see the phase diagrams in][]{Fischer18,Bastian:2021}, the energy density jumps between the onsets of quark matter and reaching the pure quark matter phase increase simultaneously. 
Consequently, most RDF EOS have significantly larger latent heat for the finite entropy configurations than at $T=0$.
This is one of the key aspects of having a strong first-order phase transition in the CCSN simulations, which features the formation of a strong hydrodynamical shock, as will be discussed further below.
All RDF models have large latent heat, especially at finite entropy, which is one particular feature of this EOS class.
However, RDF-1.3--1.7 have somewhat higher onset densities and smaller maximum masses than RDF-1.1, RDF-1.2, RDF-1.8, and RDF-1.9, such that no stable remnants could be found for any of the CCSN simulations performed in this study for any of the RDF-1.3--1.7, as will be discussed further below.

RDF EOS sets with a particularly low onset density are the ones RDF-1.8 and RDF-1.9, which are achieved due to a low linear vector coupling for RDF-1.8 as well as lacking higher-order vector repulsion for RDF-1.9 \citep[further details can be found in][]{Bastian:2021}. From these data, it becomes evident that there is a strong temperature dependence on the phase transition, in particular on the onset density. Note the temperatures for the different EOS reaching values on the order of $40$--$60$~MeV. Especially for the two RDF EOS with a low onset density at $T=0$, RDF-1.8, and RDF-1.9, with onset density on the order of nuclear saturation density for $s=3~k_{\rm B}$. Table~\ref{tab:EOS} also lists the maximum masses for all EOSs, from which it becomes evident that the effect of finite temperature enhances the maximum masses slightly, with only a few exceptions, for the entropy explored here as a representative value.
This is in agreement with what has been reported previously, also based on the simplistic thermodynamic bag model EOS \citep[c.f.][and references therein]{Khosravi:2022}.

\begin{figure*}[htp]
\begin{center}
\subfigure[~Post-bounce times corresponding to the onset of the phase transition and PNS collapse, $t_{\rm PT}$ and $t_{\rm collapse}$]{
\includegraphics[angle=0.,width=1.5\columnwidth]{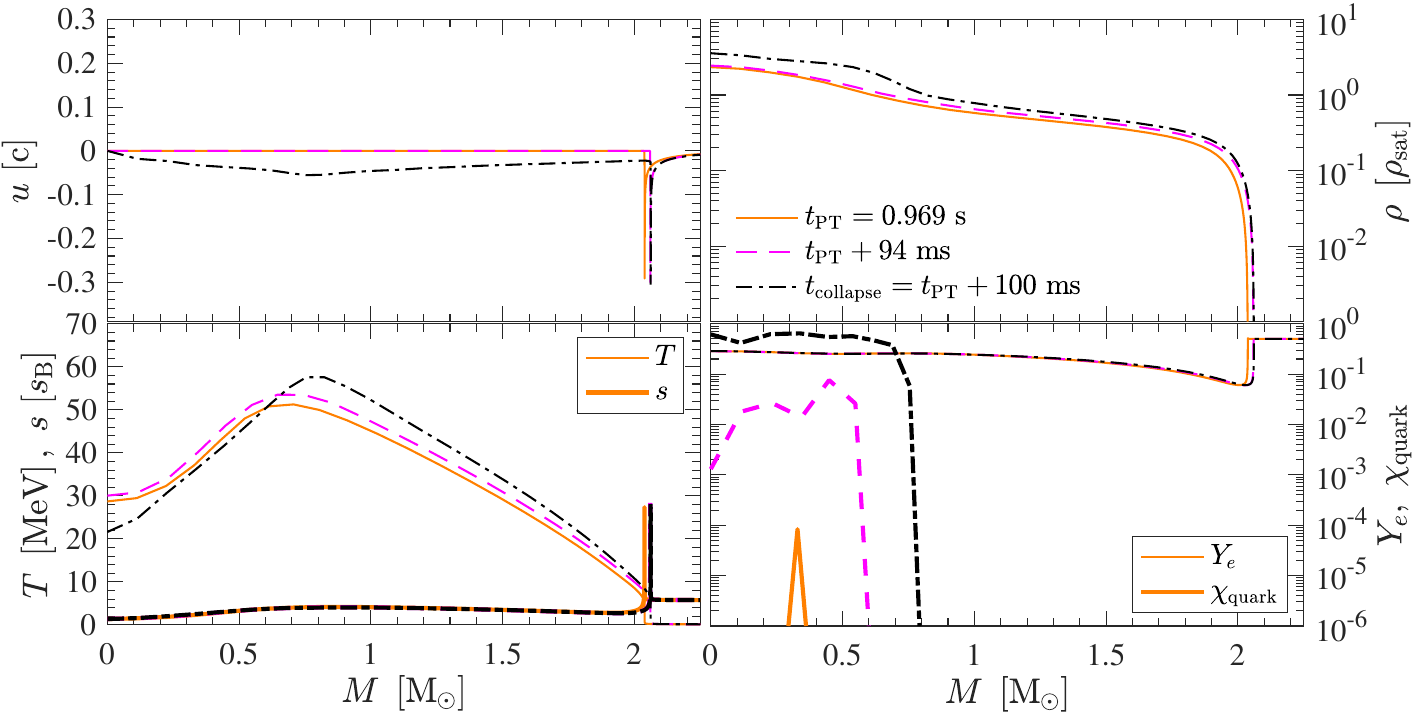}
\label{fig:quarkstate_aa}}
\subfigure[~Formation of the second shock]{
\includegraphics[angle=0.,width=1.5\columnwidth]{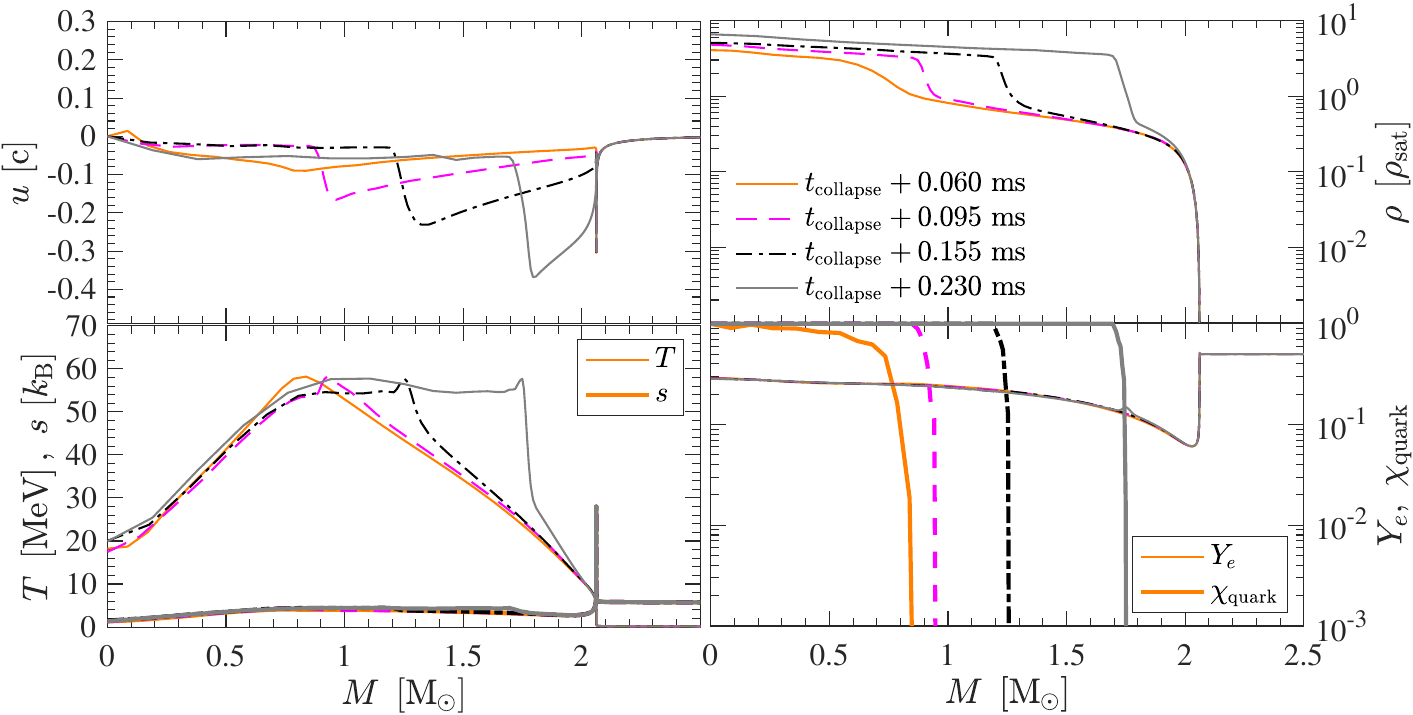}
\label{fig:quarkstate_ab}}
\subfigure[~Moment of shock breakout]{
\includegraphics[angle=0.,width=1.5\columnwidth]{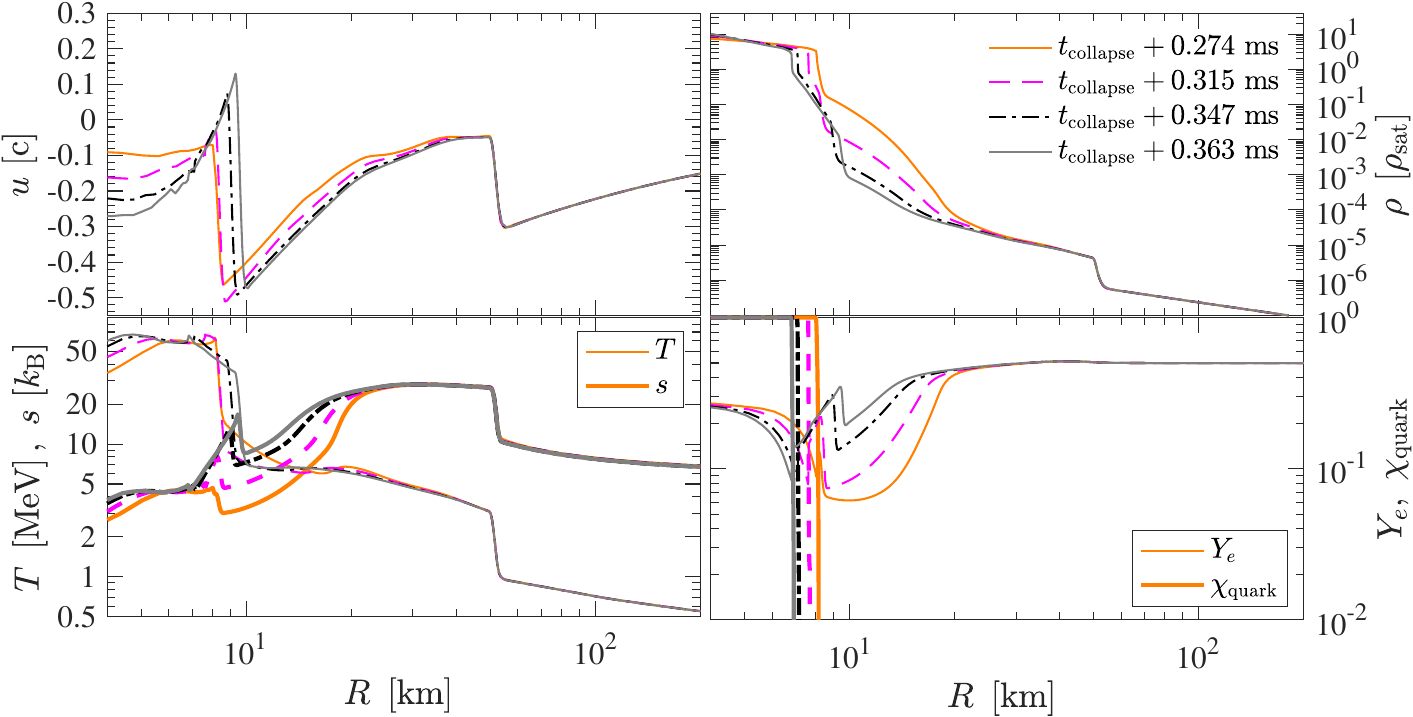}
\label{fig:quarkstate_ac}}
%
\caption{Radial profiles of selected quantities for the SN simulation of {\tt s40.0} using the RDF-1.5 EOS, showing the velocity $v$ in units of the speed of light ${\rm c}$, the rest-mass density $\rho$ in units of the saturation density $\rho_{\rm sat}$, temperature $T$ (thin lines) and entropy per particle $s$ (thick lines) as well as the electron abundance $Y_e$ (thin lines) and the quark volume fraction $\chi_{\rm quark}$ (thick lines) with respect to the enclosed baryon mass $M$ for the onset of the PNS collapse in graph~(a) and for the shock formation in graph~(b), and as a function of the radius $R$ for the phase of shock breakout in graph~(c), at selected post-bounce times (see the text and Table~\ref{tab:results} for the definitions of $t_{\rm PT}$ and $t_{\rm collapse}$).}
\label{fig:quarkstate_a}
\end{center}
\end{figure*}

Relevant for the SN applications is not only the dependence on temperature but also on $Y_e$. The latter is modulated within the string-flip model due to the inclusion of the $\rho$-meson equivalent term, for which the parameters are chosen to minimize the dependence on $Y_e$. This has been discussed in great detail in \citet{Bastian:2021}. A consequence is, e.g., that the phase boundaries for the onset of the phase transition are mildly dependent on the value of $Y_e$. 

Weak interactions in quark matter must be considered. Therefore,  baryonic degrees of freedom are reconstructed, using the following relations between up- and down-quark chemical potentials, ($\mu_{\rm u}$) and ($\mu_{\rm d}$), respectively, and the baryon and charge chemical potentials,
\begin{eqnarray}
\mu_{\rm B} = \mu_{\rm u} + 2\mu_{\rm d}~,\qquad \mu_{\rm Q} = \mu_{\rm u} - \mu_{\rm d}~.
\end{eqnarray}
This approximation is justified here since neutrinos in the quark matter phase are completely trapped for the entire simulation times considered in the present study. 
It has been applied in all previous SN studies of quark matter \citep[][]{Sagert09,Fischer11,Zha20,Fischer:2021,Kuroda2022}. 
Only after $\mathcal{O}(10~{\rm s})$ after the SN explosion onset, neutrinos will begin to decouple in the quark matter phase \citep[c.f.][]{Fischer20}, after which this simplification can no longer be applied and neutrino opacities in quark matter with QCD degrees of freedom must be employed for any prediction of the PNS evolution, in particular, for the neutrino signal and the associated deleptonization.  

\subsection{Evolutionary Scenario -- A Representative Case}
The canonical post-bounce evolution of CCSN simulations in spherical symmetry features an extended mass accretion period, which lasts for several hundreds of milliseconds. Material falls through the standing shock wave being heated and accumulates onto the central PNS. Consequently, the gravitational potential steepens, and the PNS contracts, leading to the continuous rise of the central density and temperature. In spherically symmetric simulations, this phase would continue until the maximum mass, given by the EOS, is reached, which then yields the collapse of the PNS and the formation of a black hole~\citep[c.f.][and references therein]{Sumiyoshi06,Fischer09,O'Connor11}.

However, the phase transition from normal nuclear matter to quark matter at the PNS interior terminates this evolution toward direct black hole formation.
The evolution is illustrated in Fig.~\ref{fig:quarkstate_a} as an example of a 40~M$_\odot$ progenitor model from the stellar evolution series of \citet{Woosley:2002zz}, henceforth denoted as {\tt s40.0}, using the RDF-1.5 EOS.
The moment for the onset of the phase transition, $t_{\rm PT}$, the very central fluid element is in the mixed phase, when the quark matter volume fraction becomes finite, defined when $\chi>10^{-10}$.
At $t_{\rm PT}$, the central quark matter volume fraction $\chi_{\rm quark}$ starts to rise. 
However, even though the PNS enters the mixed phase, it does not become gravitationally unstable {at this moment}, despite the substantially reduced pressure gradient. 
Only when the mass enclosed in the mixed phase reaches (or exceeds) about 0.5~M$_\odot$, the central PNS collapses, as indicated by the radial velocity profiles in Fig.~\ref{fig:quarkstate_aa} (upper left panel), which takes about 100~ms after $t_{\rm PT}$.
This is the canonical situation we observe in all our simulations. 
Once the PNS collapses, indicted by $t_{\rm collapse}$, which is defined as when the PNS infall velocities reach a few hundred kilometers per second, 
it proceeds supersonically on a sub-millisecond timescale, reaching infall velocities on the order of 10--20\% of the speed of light. 
Since the collapse proceeds adiabatically---entropy per particle and $Y_e$ remain constant---central temperature and density rise. 
The PNS compresses through the hadron-quark mixed phase, and the quark volume fraction rises consequently. 
Once reaching the pure quark matter phase at higher densities (see the upper right panel of Fig.~\ref{fig:quarkstate_ab}), the pressure slope steepens and the collapse halts with the formation of a strong, second hydrodynamics shock wave.
After that, this shock wave propagates outward while material still falls onto this shock wave, leading to the continuous growth of the quark matter core and the steepening of the velocity gradient of the shock wave, reaching velocities on the order of $50\%$ of the speed of light and peak temperatures beyond 60~MeV (see the bottom left panel in Fig.~\ref{fig:quarkstate_ab}). 

Once the second shock wave reaches the PNS surface, where the density decreases rapidly by several orders of magnitude, the mass accretion onto the second shock drops. 
Consequently, the second shock wave expands further, reaching positive matter outflow velocities, even on the order of several tenths of the speed of light (upper left panel in Fig.~\ref{fig:quarkstate_ac}).
During all this evolution through the PNS collapse, shock formation, and expansion, the first, bounce shock remains unaffected at a radius of about 80~km. 
The subsequent evolution leading to shock breakout is defined by the occurrence of positive matter velocities in the post-shock layer for the first time after the formation of the second shock wave. The onset or failure of a SN explosion depends on the equation of state for the hadron-quark hybrid EOS. 
If the accumulated PNS mass, now with the quark matter core, exceeds the maximum mass of the {hadron-quark hybrid} EOS, the successful shock expansion will be terminated by the proto-neutron star collapse and the formation of a black hole. 
This is the case for the 25~M$_\odot$ progenitor discussed here. 
Matter infall velocities behind the expanding second shock become larger and larger as the simulation proceeds (upper left panel in Fig.~\ref{fig:quarkstate_ab}). 
We continue the simulation until the central lapse function, $\alpha$, which is the $00$-component of the metric tensor and determines for instance the time dilatation, diverges to numerical values of about $10^{-2}$, beyond which stable solutions of the radiation hydrodynamics equations cannot be obtained anymore within the co-moving coordinates of {\tt AGILE-BOLTZTRAN}.

\section{Stellar model dependence}
\label{sec:prog}
SN simulations are launched from massive progenitors with ZAMS masses in the range of 25--40~M$_\odot$. This progenitor mass range has been suggested in previous studies as potentially most likely for the QCD phase transition to occur due to the generally more compact PNS achieved in these SN simulations, with higher central densities compared to lighter progenitors. We investigate the stellar models {\tt s25a28}, {\tt s30a28}, and {\tt s40a28} from \citet{Rauscher:2002} and the model {\tt s40.0} from \citet{Woosley:2002zz}. The former differ by the implementation of updated nuclear reaction rates, revised opacity tables, neutrino losses, and weak interaction rates. 

\begin{figure*}[t!]
\begin{center}
\subfigure[]{\includegraphics[angle=0.0,width=1.015\columnwidth]{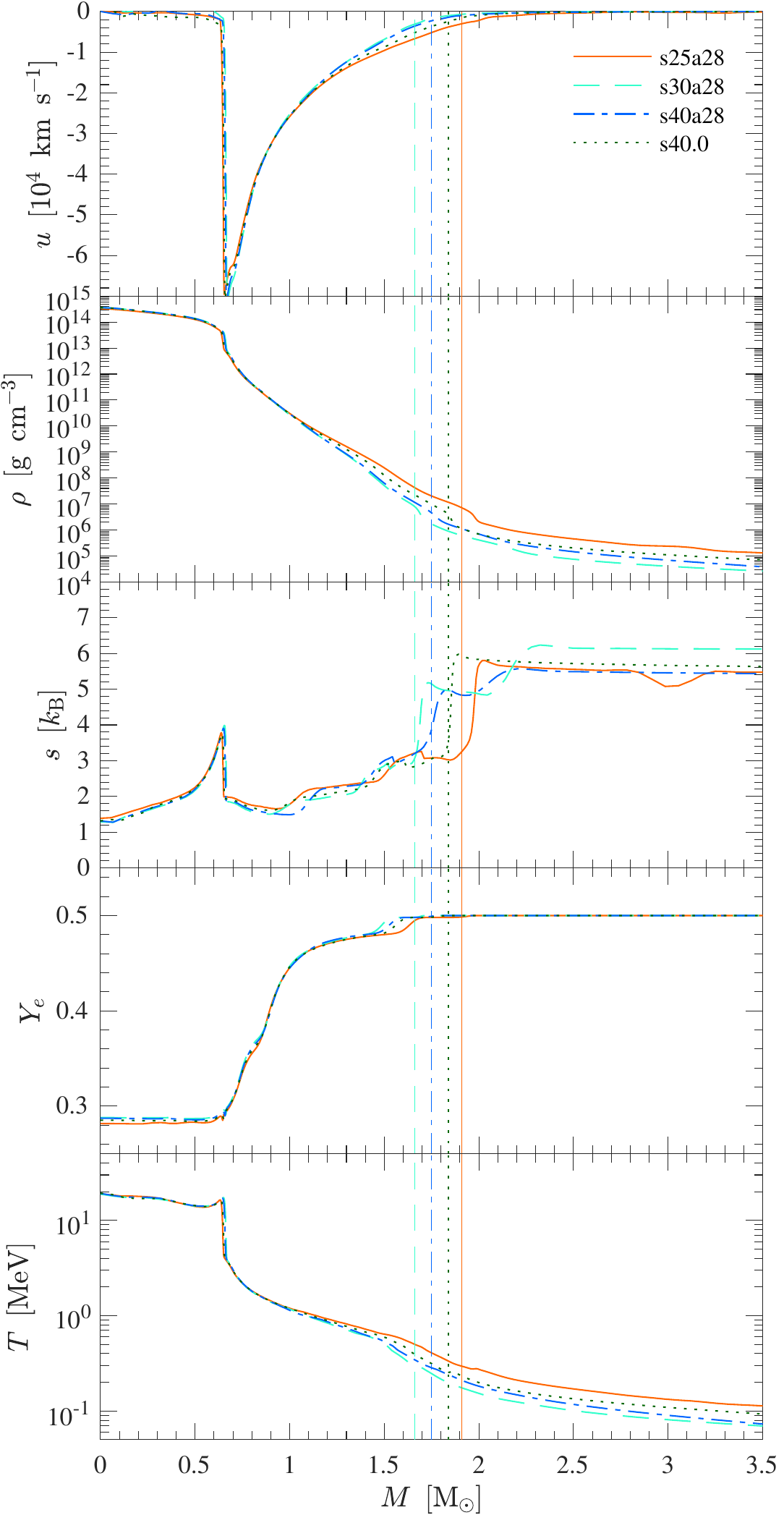}\label{fig:progs_fullstate}}
\subfigure[]{\includegraphics[angle=0.0,width=0.99\columnwidth]{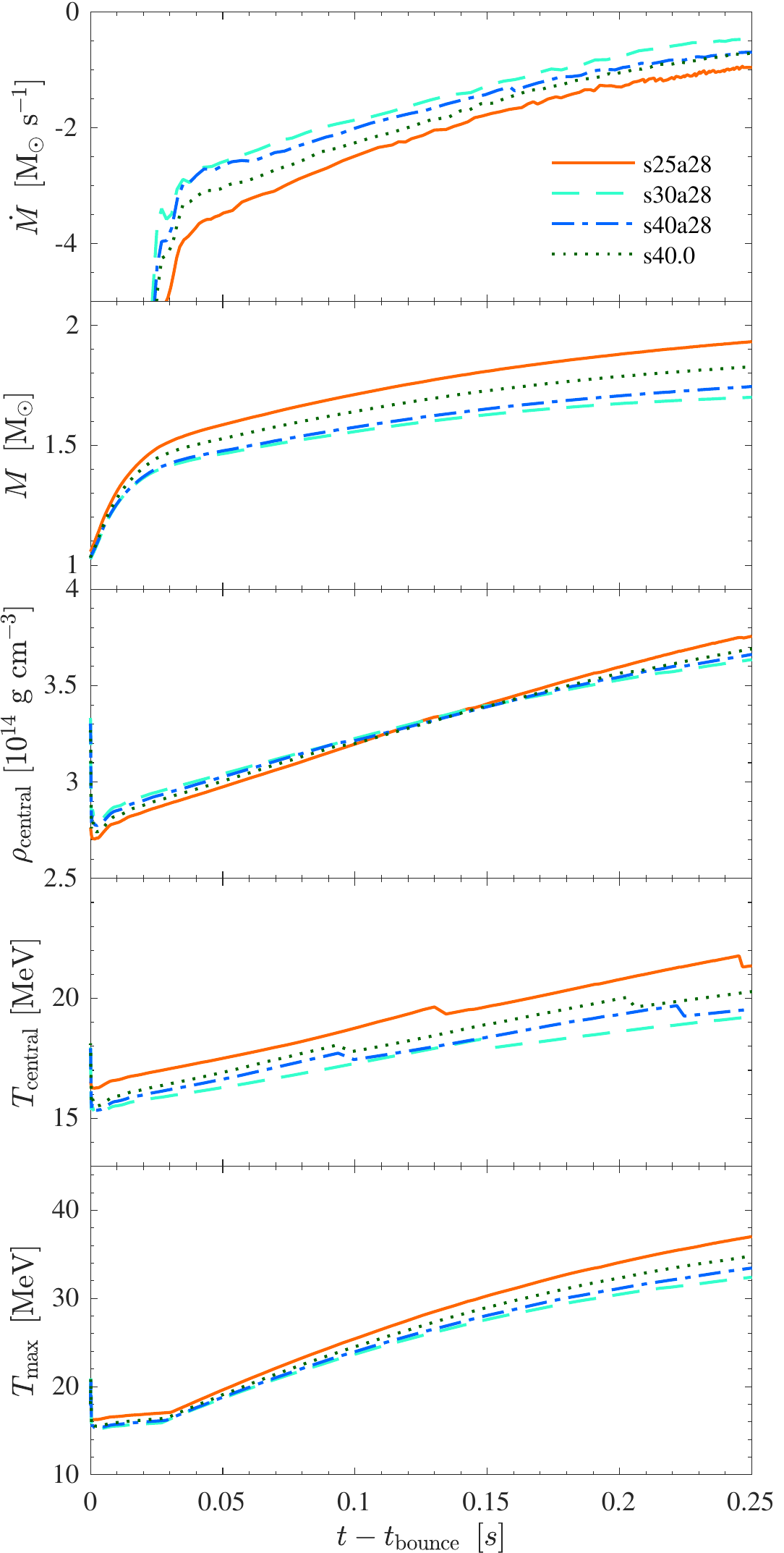}\label{fig:progs_evol}}
\caption{{\em Graph~(a):}~Radial bounce profiles as a function of the enclosed baryon mass for the different progenitors under investigation, from top to bottom showing the radial velocity $u$, restmass density $\rho$, entropy per baryon $s$, electron abundance $Y_e$ and temperature $T$. {\em Graph~(b):} Post-bounce evolution of selected quantities, showing from top to bottom the mass accretion rate $\dot{M}$ (see the text for the definition), the enclosed baryon mass of the PNS $M_{\rm PNS}$, both evaluated at a density of $\rho=10^{11}$~g~cm$^{-3}$, the central density $\rho_{\rm central}$ as well as central and maximum temperatures, $T_{\rm central}$ and $T_{\rm max}$, respectively.}
\label{fig:progs}
\end{center}
\end{figure*}

\begin{table}[htp]
\begin{center}
\caption{Selected Properties of the Stellar Progenitor Models.}
\begin{tabular}{lcccc}
\hline
\hline
 Progenitor & $M_{\rm Fe}^a$ & $\rho_{\rm Fe-Si}^b$ & $M_{\rm Fe-Si}^c$ & $\rho_{\rm Fe-Si}^d$  \\
& $[$M$_\odot]$ & $[10^7\,\,\,{\rm g\,\,\,cm}^{-3}]$ & $[$M$_\odot]$ & $[10^6\,\,\,{\rm g\,\,\,cm}^{-3}]$ \\
\hline
{\tt s25a28} & 1.63 & 2.1 & 1.91 & 5.1 \\
{\tt s30a28} & 1.48 & 2.6 & 1.66 & 5.8 \\
{\tt s40a28} & 1.52 & 2.2 & 1.75 & 4.1 \\
{\tt s40.0} & 1.56 & 2.2 & 1.84 & 3.5 \\
\hline
\end{tabular}
\end{center}
{\bf Notes.}\\
$^a$~Iron-core masses \\
$^b$~Density at the iron-core silicon-sulfur layer transition \\
$^c$~Enclosed mass at the iron-core silicon-sulfur layer transition  \\
$^d$~Corresponding density at the iron-core silicon-sulfur layer transition \\
\label{tab:prog}
\end{table}

Progenitor differences, relevant for the post-bounce SN evolution, arise at the transitions from the iron cores to the silicon-sulfur layers, defining the iron-core masses. 
These coincide with the NSE to non-NSE transitions in the SN simulations. 
The iron-core masses for all progenitor models under investigation are listed in Table~\ref{tab:prog}, together with the corresponding densities. 
Figure~\ref{fig:progs_fullstate} compares all stellar models at core bounce, which we define when the maximum central density is reached,
corresponding to the moment just before the shock breakout when positive matter velocities are obtained for the first time after core bounce 
(see the velocities $u$ at the top panel).
 The SN simulation results illustrated in Fig.~\ref{fig:progs} are all performed with the RDF-1.8 EOS, which at densities below the onset of quark matter is based on the DD2 hadronic EOS.
Note that at core bounce quark matter has not appeared in any of the simulations. However, there are slight differences, which are due to the mismatch in the temporal sampling of the SN simulation data.
 Small differences in the core structure, c.f. the entropy and $Y_e$ profiles in Fig.~\ref{fig:progs_fullstate}, are related to the different Fe-core masses at the onset of stellar core collapse. 

Larger differences are found toward the outer layers of the inner stellar cores, where the entropies rise sharply, at a mass coordinate of about 1.55--1.95~M$_\odot$ (see the dashed vertical lines in Fig.~\ref{fig:progs_fullstate} with the matching color of each progenitor model), which corresponds to the stellar core's silicon and sulfur abundances dropping largely. 
Note that these exclude the oxygen-burning shells of the stellar cores. 
The associated mass coordinates are denoted as $M_{\rm Fe-Si}$ and are listed in Table~\ref{tab:prog} for all progenitors under investigation. 
Note further that in stellar modeling, this aligns with the transition from radiation to convection-dominated regimes. 

The structure of the layers above the iron core determines the post-bounce evolution. 
Simulations featuring a high mass accretion rate correspond to progenitor models with a high rest-mass density in the layers above the iron core. 
Particularly relevant are the different enclosed masses at the inner-to-outer core transitions, i.e. at $M_{Fe-Si}$ (see Table~\ref{tab:prog}), rather than the iron-core masses or the silicon-sulfur layer densities. 
Especially {\tt s25a28} features the largest core masses and hence results in the highest post-bounce mass accretion rates, $\dot{M}(R)=4\pi R^2 u(R) \rho(R)$, which are evaluated at a certain radius $R$, velocity $u$ and rest-mass density $\rho$, as illustrated in Fig.~\ref{fig:progs_evol} (top panel), during the early post-bounce evolution prior to the phase transition.
Mass accretion rates and enclosed PNS mass, $M_{\rm PNS}$, are evaluated at a rest-mass density of $\rho=10^{11}$~g~cm$^{-3}$. 
As a consequence of the highest mass accretion rates for {\tt s25a28}, the enclosed PNS mass $M_{\rm PNS}$ quickly approaches high values, on the order of 2~M$_\odot$ already after a few hundred milliseconds post-bounce, however, with the lowest central densities $\rho_{\rm central}$ due to the larger thermal support (see the maximum temperatures $T_{\rm max}$ shown in the bottom panel in Fig.~\ref{fig:progs_evol}). 
Contrary, the progenitors with the lowest mass accretion rates, {\tt s30a28} and {\tt s40a28}, have the slowest PNS mass growth and hence the highest central densities during the post-bounce evolution as well as the lowest $T_{\rm max}$. 
Hence, such stellar models are most favorable to yield stable remnants after a possible hadron-quark phase transition, in particular, if the enclosed PNS mass is below the EOS critical mass at the phase transition onset. 

It is interesting to note that despite the same ZAMS mass and similar iron-core properties for the 40~M$_\odot$ progenitors, {\tt s40a28} and {\tt s40.0}, the different implementation of nuclear rates and mass loss gives rise to different structures in terms of the rest-mass density and the entropy profiles in the silicon-sulfur layers. 
This has consequences on the QCD phase transition in SN simulations, which will be discussed further below.

\section{Systematics of the Hadron-Quark Phase Transition}
\label{sec:systematics}
In order to obtain a systematic understanding of the potential impact of the hadron-quark phase transition on SN phenomenology, simulations are discussed that are launched from the progenitors introduced in sec.~\ref{sec:prog}, for all nine RDF EOSs under investigation. 
Table~\ref{tab:results} below contains a summary of selected quantities from all these runs. Note that SN simulations for {\tt s35a28} of \citet{Rauscher:2002} are performed too, however, found no quantitative differences to the one of {\tt s30a28}. 
Both progenitors' stellar core structures are nearly identical. Quantitative results obtained with respect to the hadron-quark phase transition and the subsequent SN evolution for {\tt s30a28} apply equally to {\tt s35a28}. 

\begin{figure}[htp]
\begin{center}
\subfigure[~Explosion model ({\tt s30a28} RDF-1.2)]{\includegraphics[angle=0.,width=0.875\columnwidth]{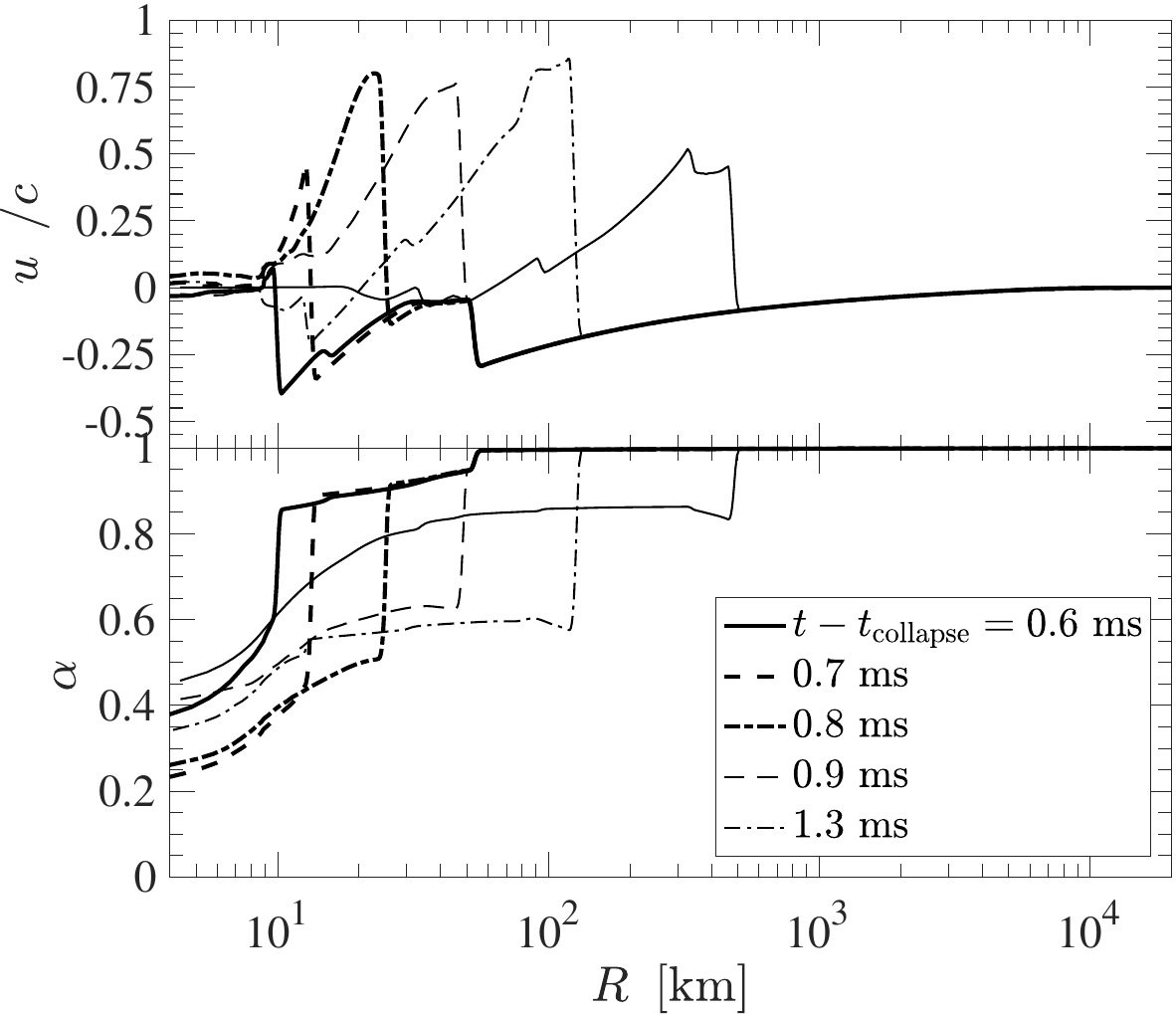}\label{fig:results_a}}
\subfigure[~Black hole formation (prompt collapse, {\tt s25a28} RDF-1.1)]{\includegraphics[angle=0.,width=0.875\columnwidth]{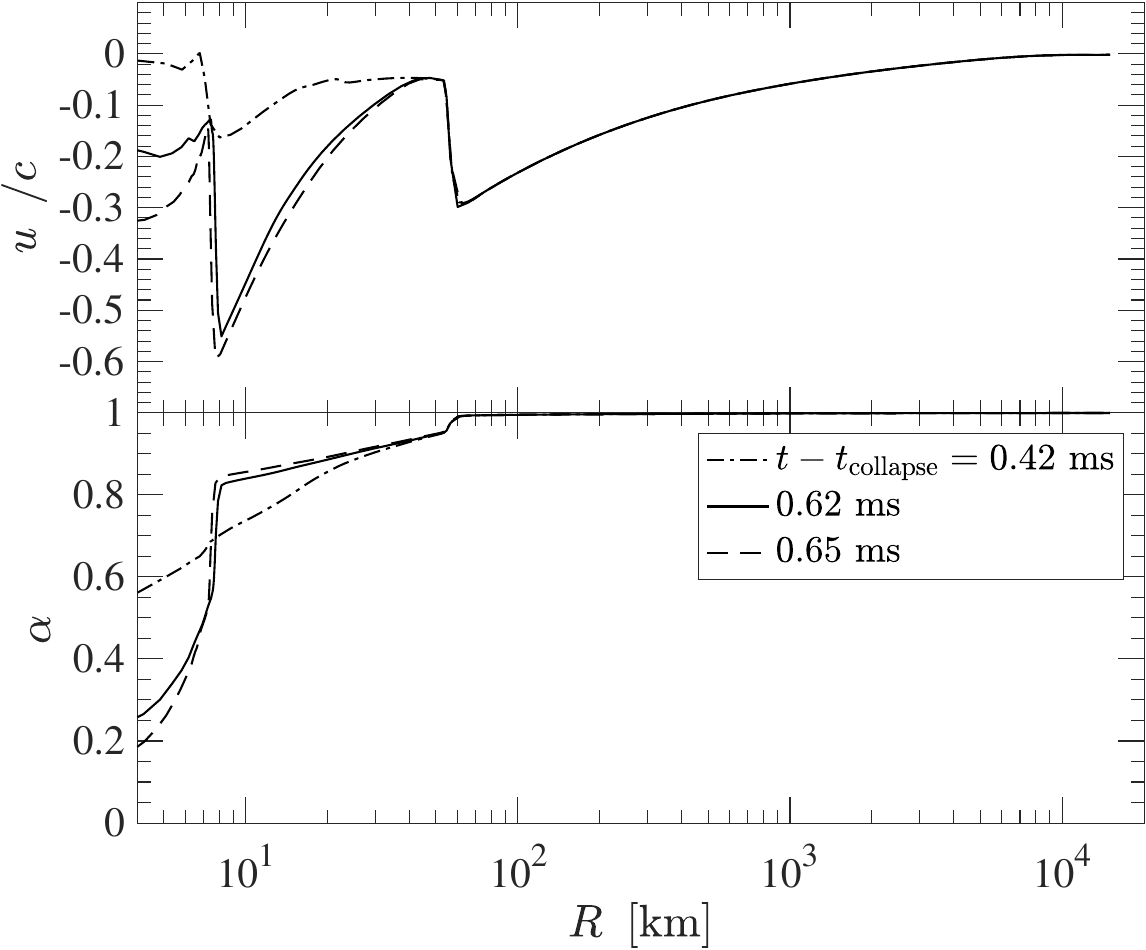}\label{fig:results_b}}
\subfigure[~Black hole formation (delayed collapse, {\tt s30a28} RDF-1.6)]{\includegraphics[angle=0.,width=0.875\columnwidth]{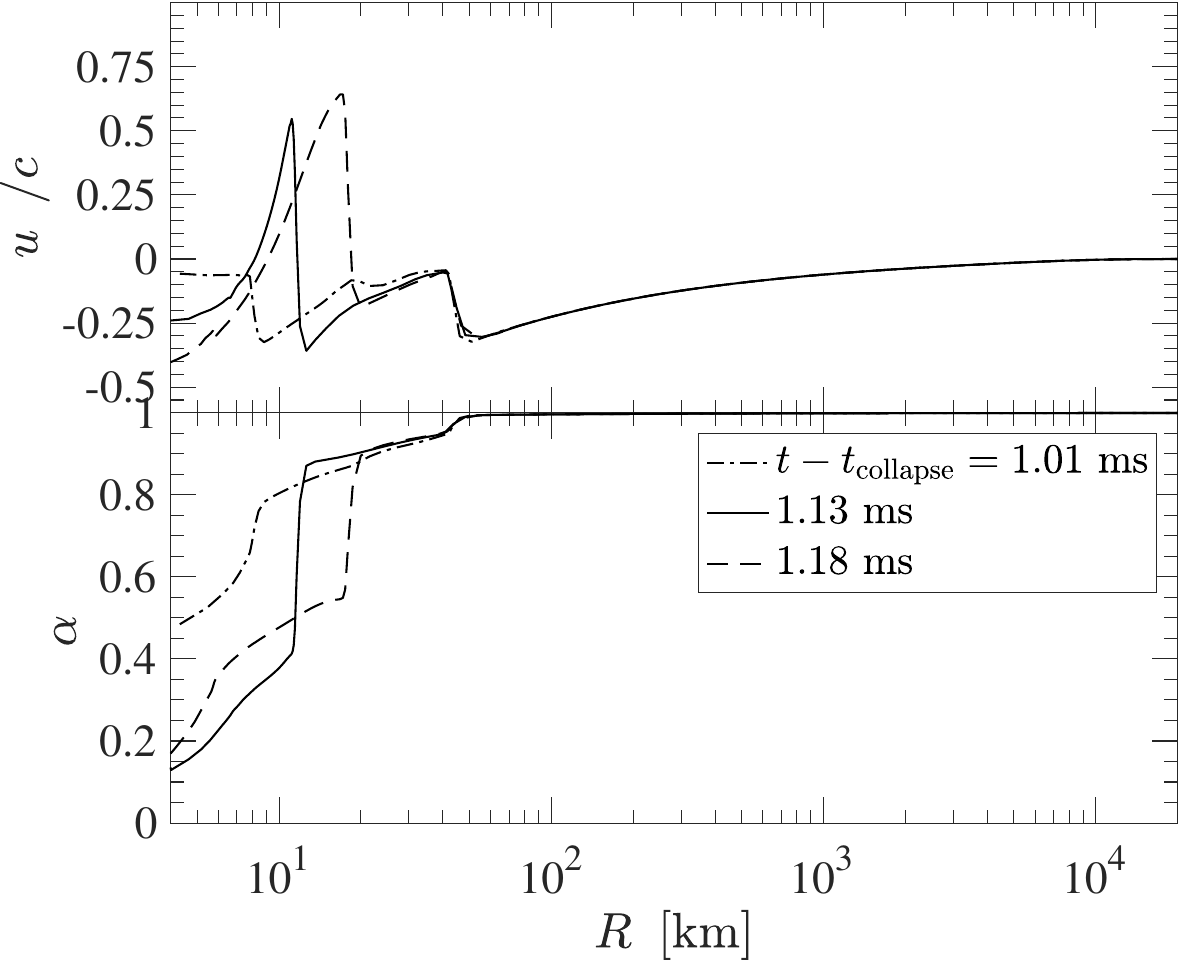}\label{fig:results_c}}
\caption{Early shock evolution for a representative explosion model (top panel) and two failed models, separated into prompt black hole formation (middle panel) and delayed (bottom panel), showing radial profiles of the velocity $u$, in units of the speed of light $c$, and lapse function $\alpha$.}
\label{fig:results}
\end{center}
\end{figure}

Simulations featuring a high mass accretion rate quickly reach the phase transition onset densities during the post-bounce evolution. 
In particular, {\tt s25a28} has the highest post-bounce mass accretion rate and hence reaches the onset conditions earliest for all EOSs, see $t_{\rm PT}$ in Table~\ref{tab:results}, by several hundreds of milliseconds in comparison to {\tt s30a28}, {\tt s40a28} and {\tt s40.0}. 
Related, the enclosed baryon masses are higher at the moment of PNS collapse, $M_{\rm collapse}$, for {\tt s25s28} in comparison to all other stellar progenitors. 
For all runs launched from {\tt s25a28}, the enclosed mass exceeds the maximum mass of the RDF EOS at the onset of the PNS collapse. 
Hence, the remnants collapse into black holes. 
The only exception is RDF-1.9, for which the phase transition onset occurs as early as 226~ms after the core bounce, with the enclosed mass below the maximum mass of RDF-1.9 (see Table~\ref{tab:results}). 

\begin{figure}[htp]
\begin{center}
\subfigure[]{
\includegraphics[angle=0.,width=1.\columnwidth]{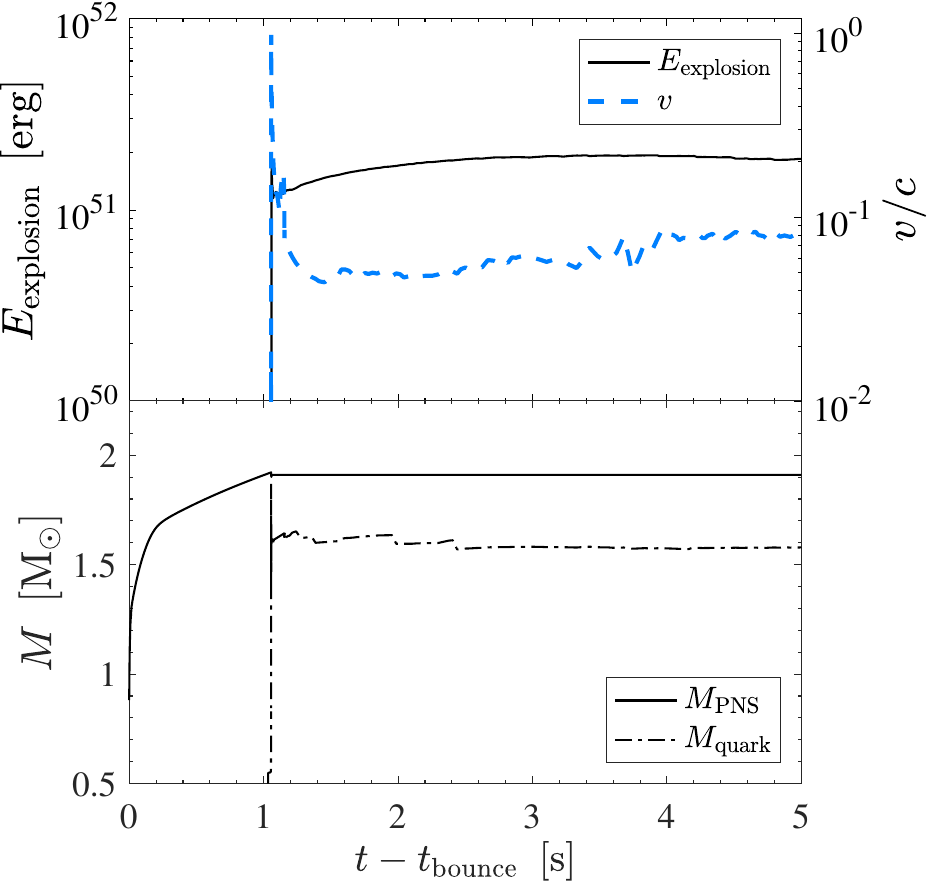}\label{fig:s30_expl}}
\\
\subfigure[]{\includegraphics[angle=0.,width=1.\columnwidth]{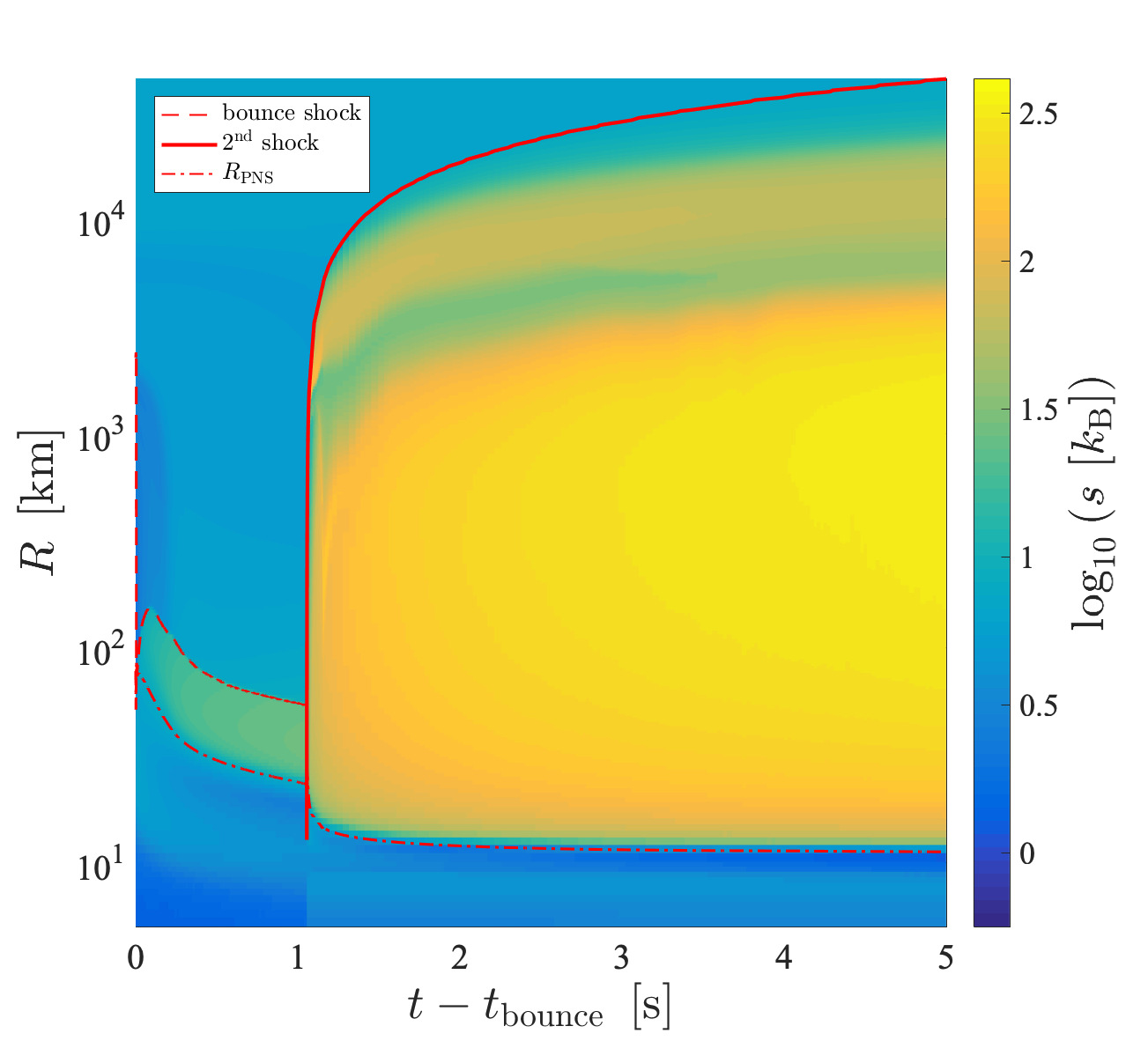}\label{fig:s30_shellplot}}
\caption{~Post-bounce evolution of the RDF-1.2 {\tt s30a28} explosion model. {\em Graph~(a):} ~Diagnostic explosion energy $E_{\rm explosion}$ and the normalized maximum velocity of the shock $v$ (top panel) as well as the enclosed PNS mass $M_{\rm PNS}$ and the quark core mass $M_{\rm quark}$ (bottom panel). {\em Graph~(b):}~Space-time diagram up to 5~seconds with the entropy color coded showing the shock locations of the bounce shock (red dashed line) and the second shock (red solid line) as well as the PNS radius (red dashed-dotted line) defined as $R(\rho=10^{11}$~g~cm$^{-3})$.}
\label{fig:s30_explosion}
\end{center}
\end{figure}

In the following analysis, we distinguish between exploding and failed models. 
As an example of the former, Fig.~\ref{fig:results_a} shows the radial profiles of the velocity at selected post-collapse times for {\tt s30a28} RDF-1.2. 
All other exploding models qualitatively agree with the evolution of this one. 
The evolution shown in Fig.~\ref{fig:results_a} corresponds to the shock breakout. 
The latter is characterized by the rapid shock expansion, on the order of a few milliseconds, to increasingly larger radii. 
Thereby, taking over and merging with the standing bounce shock, which is illustrated in the space-time diagram, Fig.~\ref{fig:s30_shellplot}, showing the evolution of the bounce shock (red dashed line) as well as the formation and propagation of the second shock (red solid line),
reaching relativistic velocities on the order of the speed of light. 
Furthermore, the shock expands to radii on the order of several $10^4$~km on a timescale of a few hundred milliseconds. 
Note that initially, massive quark cores form on the order of $M_{\rm quark}\simeq$1.5--1.8~M$_\odot$  (see the bottom panel in Fig.~\ref{fig:s30_expl}). 
The latter is defined as the baryonic mass contained inside the PNS that has $\chi_{\rm quark}=1$. These initially massive quark matter cores are due to the high temperatures reached at the PNS interiors on the order of $50$--$60$~MeV.
During the later evolution of the PNS deleptonization, on the order of several tens of seconds, the core temperatures decrease, and $M_{\rm quark}$ will reduce.
After around 20~s in this model, we find $M_{\rm quark}\simeq 1.06$~M$_\odot$, still decreasing, with a central temperature of around 41~MeV ($s\simeq 3.0~k_{\rm B}$) and a central lepton fraction of $Y_{\rm L}=0.018$. 
This demonstrates the complexity of the system and the difficulties in relating the findings in dynamical simulations with those obtained in static solutions, even for finite entropy configurations and taking approximately the role of a finite neutrino abundance into account.

Values of the diagnostic explosion energy, $E_{\rm explosion}$, are listed in Table~\ref{tab:fit} for all explosion runs. 
These are computed following the standard procedure \citep[c.f.][and references therein]{Fischer10}, where the radial profiles of the total specific energy are integrated from the stellar surface toward the center.
Note that here $E_{\rm explosion}$ contains the contributions from the stellar progenitor envelopes. 
Therefore, the stellar envelopes are matched carefully with the SN simulation domains, which are considered up to on the order of $10^5$~km for all progenitors under investigation. 
The values for the diagnostic explosion energy are obtained at asymptotically late times on the order of a few seconds after the explosion onset, however, excluding the later evolution of the PNS deleptonization. 

\begin{table*}[htp]
\caption{Summary of the DD2/DD2F/DD2Fev-RDF EOS SN Simulations.}
\begin{tabularx}{0.99\textwidth}{lr@{\hspace{1cm}}c@{\hspace{1cm}}c@{\hspace{1cm}}c@{\hspace{0.7cm}}c@{\hspace{0.7cm}}c@{\hspace{0.7cm}}c}
\hline
\hline
Progenitor & EOS & $t_{\rm PT}^a$ & $t_{\rm collapse}^b$ & $M_{\rm collapse}^c$ & $\rho_{\rm collapse}^d$ & $\triangle t_{\rm breakout}^e$ & $M_{\rm remnant}^f$  \\
 & hadronic \; quark & $[{\rm s}]$ & $[{\rm s}]$ & $[{\rm M}_\odot]$ & $[10^{14}\,\,{\rm g\,\,\,cm}^{-3}]$ & $[{\rm ms}]$ & $[{\rm M}_\odot]$  \\
 \hline
{\tt s25a28}$^{\dagger}$ & DD2F \; RDF-1.1 & 0.780 & 0.879 & 2.22 & 6.08 & 2.16 & 2.22  \\ 
{\tt s25a28}$^{\ddagger}$ & DD2Fev \; RDF-1.2 & 0.540 & 0.621 & 2.13 & 5.18 & 4.17 & 2.13  \\  
{\tt s25a28}$^{\dagger}$ & DD2F \; RDF-1.3 & 0.782 & 0.823 & 2.20 & 6.00 & 0.94 & 2.20 \\ 
{\tt s25a28}$^{\ddagger}$ & DD2F \; RDF-1.4 & 0.845 & 0.889 & 2.22 & 6.07 & 4.89 & 2.22 \\ 
{\tt s25a28}$^{\dagger}$ & DD2F \; RDF-1.5 & 0.774 & 0.774 & 2.18 & 5.63 & 1.01 & 2.18 \\  
{\tt s25a28}$^{\dagger}$ & DD2F \; RDF-1.6 & 0.845 & 0.899 & 2.23 & 6.14 & 2.58 & 2.23 \\ 
{\tt s25a28}$^{\dagger}$ & DD2F \; RDF-1.7 & 0.688 & 0.743 & 2.17 & 5.59 & 1.59 & 2.17 \\ 
{\tt s25a28}$^{\dagger}$ & DD2 \; RDF-1.8 & 0.358 & 0.490 & 2.10 & 4.48 & 3.73 & 2.10 \\ 
{\tt s25a28}$^{*}$ & DD2 \; RDF-1.9 & 0.226 & 0.323 & 1.99 & 4.21 & 4.47 & 1.98 \\ 
\hline
\hline
{\tt s30a28}$^{\ddagger}$ & DD2F \; RDF-1.1 & 1.393 & 1.533 & 2.04 & 5.93 & 0.80 & 2.03 \\ 
{\tt s30a28}$^{*}$ & DD2Fev \; RDF-1.2 & 0.914 & 1.054 & 1.92 & 5.79 & 1.45 & 1.91 \\ 
{\tt s30a28}$^{\ddagger}$ & DD2F \; RDF-1.3 & 1.394 & 1.429 & 2.02 & 5.86 & 2.03 & 2.02 \\ 
{\tt s30a28}$^{\ddagger}$ & DD2F \; RDF-1.4 & 1.395& 1.623 & 2.06 & 6.24 & 0.85 & 2.06 \\ 
{\tt s30a28}$^{\ddagger}$ & DD2F \; RDF-1.5 & 1.211 & 1.358 & 2.00 & 5.63 & 1.93 & 2.00 \\ 
{\tt s30a28}$^{\ddagger}$ & DD2F \; RDF-1.6 & 1.394 & 1.599 & 2.05 & 5.92 & 2.01 & 2.05 \\ 
{\tt s30a28}$^{\ddagger}$ & DD2F \; RDF-1.7 & 1.184 & 1.302 & 1.99 & 2.81 & 0.75 & 1.99 \\ 
{\tt s30a28}$^{*}$ & DD2 \; RDF-1.8 & 0.700 & 0.830 & 1.87 & 4.79 & 1.75 & 1.85 \\ 
{\tt s30a28}$^{*}$ & DD2 \; RDF-1.9 & 0.354 & 0.578 & 1.81 & 4.27 & 1.34 & 1.78 \\ 
\hline
\hline
{\tt s40a28}$^{\ddagger}$ & DD2F \; RDF-1.1 & 1.181 & 1.286 & 2.08 & 6.09 & 0.95 & 2.08 \\ 
{\tt s40a28}$^{*}$ & DD2Fev \; RDF-1.2 & 0.772 & 0.893 & 1.98 & 5.68 & 1.35 & 1.96 \\  
{\tt s40a28}$^{\ddagger}$ & DD2F \; RDF-1.3 & 1.211 & 1.223 & 2.07 & 5.84 & 1.59 & 2.07 \\ 
{\tt s40a28}$^{\ddagger}$ & DD2F \; RDF-1.4 & 1.139 & 1.362 & 2.10 & 6.08 & 1.92 & 2.10 \\ 
{\tt s40a28}$^{\ddagger}$ & DD2F \; RDF-1.5 & 1.079 & 1.116 & 2.04 & 5.54 & 2.51 & 2.04 \\  
{\tt s40a28}$^{\dagger}$ & DD2F \; RDF-1.6 & 1.184 & 1.370 & 2.10 & 6.12 & 1.82 & 2.10 \\ 
{\tt s40a28}$^{\ddagger}$ & DD2F \; RDF-1.7 & 1.002 & 1.083 & 2.04 & 5.50 & 2.17 & 2.03 \\ 
{\tt s40a28}$^{*}$ & DD2\;  RDF-1.8 & 0.477 & 0.713 & 1.92 & 4.54 & 3.31 & 1.90 \\ 
{\tt s40a28}$^{*}$ & DD2 \; RDF-1.9 & 0.224 & 0.487 & 1.85 & 4.54 & 2.75 & 1.83 \\ 
\hline
\hline
{\tt s40.0}$^{\dagger}$ & DD2F \; RDF-1.1 & 1.175 & 1.208 & 2.11 & 6.30 & 2.00 & 2.10 \\ 
{\tt s40.0}$^{\ddagger}$ & DD2Fev \; RDF-1.2 & 0.626 & 0.911 & 2.02 & 5.59  & 3.30 & 2.02 \\  
{\tt s40.0}$^{\dagger}$ & DD2F \; RDF-1.3 & 1.120 & 1.181 & 2.10 & 5.82 & 1.93 & 2.10 \\ 
{\tt s40.0}$^{\dagger}$ & DD2F \; RDF-1.4 & 1.151 & 1.270 & 2.12 & 5.92 & 1.29 & 2.12 \\ 
{\tt s40.0}$^{\dagger}$ & DD2F \; RDF-1.5 & 0.969 & 1.069 & 2.07 & 5.55 & 3.95 & 2.07 \\  
{\tt s40.0}$^{\dagger}$ & DD2F \; RDF-1.6 & 1.151 & 1.270 & 2.12 & 5.92 & 1.78 & 2.12 \\ 
{\tt s40.0}$^{\ddagger}$ & DD2F \; RDF-1.7 & 0.984 & 1.030 & 2.07 & 5.54 & 2.08 & 2.06 \\ 
{\tt s40.0}$^{*}$ & DD2 \; RDF-1.8 & 0.439 & 0.690 & 1.97 & 4.50 & 1.77 & 1.95 \\ 
{\tt s40.0}$^{*}$ & DD2 \; RDF-1.9 & 0.206 & 0.438 & 1.90 & 4.50 & 3.00 & 1.88 \\ 
\hline
\hline
\end{tabularx}
\\
\\
{\bf Notes.}\\
$^a$~Post bounce time for the phase transition to occur \\
$^b$~Post bounce time for PNS collapse \\
$^c$~Enclosed baryon mass at the onset of PNS collapse \\
$^d$~Central density at the onset of PNS collapse \\
$^e$~Time delay relative to $t_{\rm collapse}$ for shock break out or black hole formation \\
$^f$~Enclosed PNS remnant baryon mass after the phase transition evaluated at a density of $10^{11}$~g~cm$^{-3}$ at the end of the supernova simulations \\ 
$^{\dagger}$~Prompt black hole formation \\
$^{\ddagger}$~Delayed black hole formation \\
$^{*}$~Explosion model \\
\label{tab:results}
\end{table*}

Representative cases of black hole formation are shown in Figs.~\ref{fig:results_a} and \ref{fig:results_b}. 
However, unlike in previous studies of failed SN associated with the QCD phase transition \citep[c.f.][]{Zha21,JakobusMueller2022}, we identify two different scenarios of black hole formation. One, prompt collapse (in Fig.~\ref{fig:results_b}) before the possible shock breakout, and two, delayed black hole formation (in Fig.~\ref{fig:results_c}). 
Table~\ref{tab:results} marks the two cases of black hole formation with different labels, in order to distinguish them. In the former case, illustrated here in the example of {\tt s25a28} with the RDF-1.1 EOS, the enclosed mass exceeds the critical mass already at the onset of the PNS collapse initiated due to the formation of a massive quark matter core. This is shown via the lapse function, $\alpha$, decreasing already below $\alpha\leq 0.2$ before positive shock velocities could be obtained and before the shock could break out from the core. 
In the case of a delayed black hole formation, illustrated for the case {\tt s30a28} RDF-1.6 (see Fig.~\ref{fig:results_c}), the second shock wave accelerates to larger radii with relativistic positive matter velocities for several tenths of a millisecond, before the PNS collapses and the black hole appears, i.e. before the lapse function decreases also for these cases $\alpha\leq 0.2$. 
Note that in several cases the second shock takes over the SN bounce shock, reaching radii on the order of more than 100~km. 

Table~\ref{tab:results} lists the post-bounce times for the onset of the QCD phase transition $t_{\rm PT}$ and the onset of the PNS collapse $t_{\rm collapse}$, which is delayed by $\sim$100~ms for some EOSs. Further listed are the enclosed PNS mass at the onset of collapse $M_{\rm collapse}$ and the corresponding central density $\rho_{\rm collapse}$ as well as the time delay for shock break out $\triangle t_{\rm breakout}$ together with the remnant masses $M_{\rm remnant}$, determined at the moment of black hole formation for the failed models and at asymptotically late times on the order of several seconds after the explosion onset for the exploding runs. Note that in the case of failed explosions, $\triangle t_{\rm breakout}$ is the time delay between $t_{\rm collapse}$ and the black hole formation, irrespective of the two aforementioned scenarios.

Furthermore, as a representative case for an exploding model, Fig.~\ref{fig:s30_expl} shows the evolution of the diagnostic explosion energy $E_{\rm explosion}$ and the enclosed PNS mass $M_{\rm PNS}$ as well as the quark core mass $M_{\rm quark}$.
Fig.~\ref{fig:s30_shellplot} shows the entire evolution for this run, from core bounce up to 5~s, with the entropy per particle color coded. Marked also are the shock locations, bounce shock, and second shock, as well as the PNS surface, defined when $\rho=10^{11}$~g~cm$^{-3}$.

It is interesting to note that SN simulations launched from {\tt s40.0} behave qualitatively similarly to those launched from {\tt s40a28}. However, only RDF-1.8 and RDF-1.9 result in explosions, while RDF-1.2 leads to an explosion for {\tt s40a28}, it belongs to the failed branch for {\tt s40.0}, precisely, to the delayed scenario in which the expanding second shock does take over the bounce shock. The reason for the different evolution between {\tt s40.0} and {\tt s40a28} is the larger enclosed mass and the somewhat faster growth of the PNS mass, as is illustrated in Fig.~\ref{fig:progs_evol}. This is caused by a higher late-time post-bounce mass accretion rate for {\tt s40.0} prior to the phase transition, due to a slightly higher density in the silicon-sulfur layer of the progenitor. These findings demonstrate the sensitivity of the QCD phase transition SN explosion mechanism on the stellar progenitor. 

\begin{figure*}[htp]
\subfigure[~{\tt s30a28}]{\includegraphics[angle=0.,width=0.975\columnwidth]{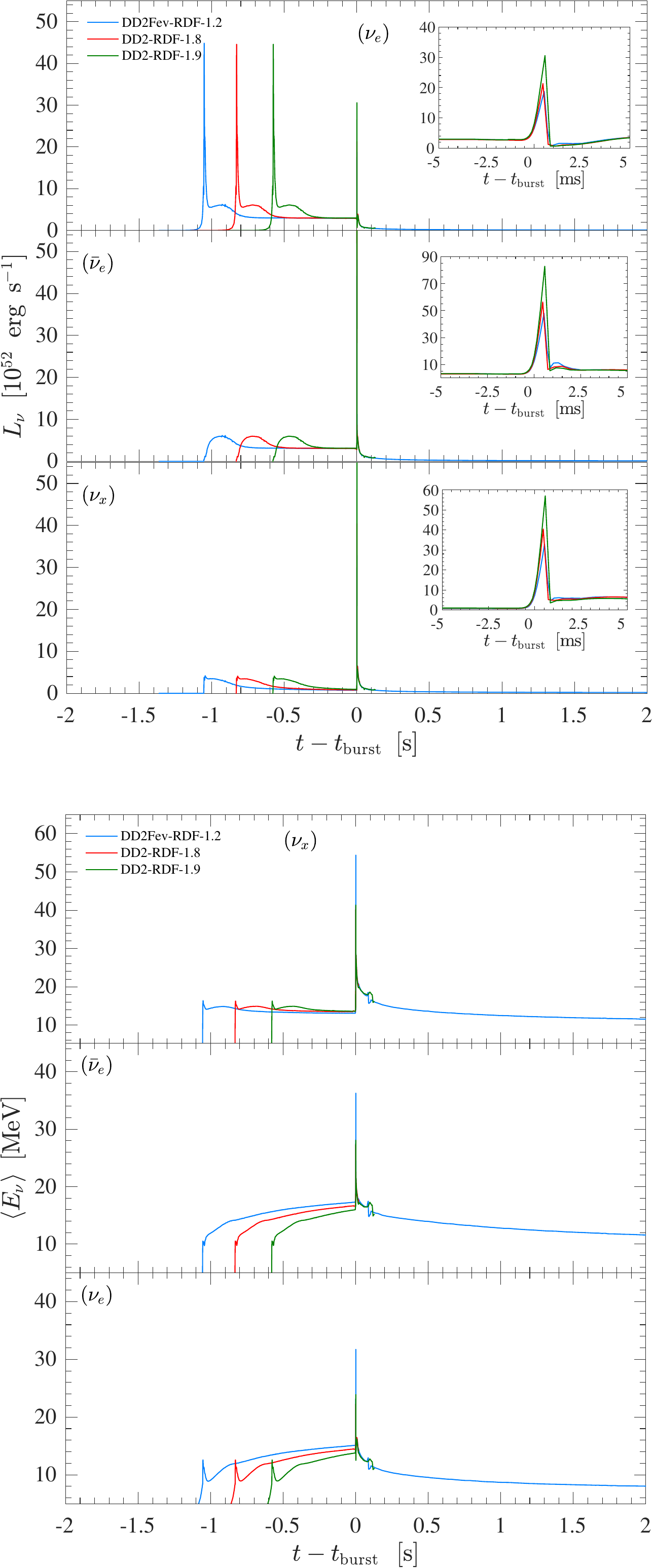}
\label{fig:lumin_s30a28}}
\subfigure[~{\tt s40a28}]{\includegraphics[angle=0.,width=0.965\columnwidth]{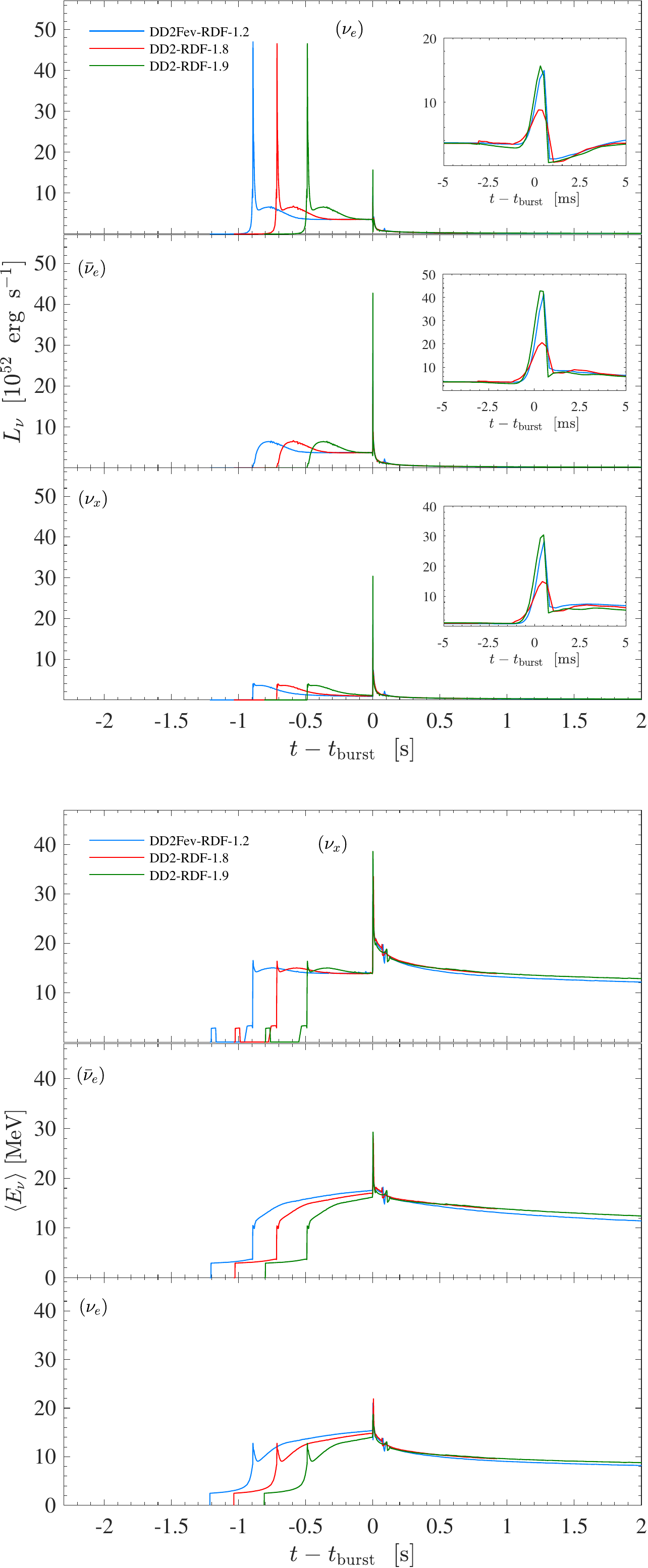}}
\label{fig:lumin_s40a28}
\caption{Neutrino luminosity $L_\nu$ (top panels) and average energy $\langle E_\nu \rangle$ (bottom panels) evolution with respect to the post-bounce time of the release of the second neutrino burst $t_{\rm burst}$, for the explosion models of {\tt s30a28} in graph~(a) and {\tt s40a28} in graph~(b), distinguishing $\nu_e$, $\bar\nu_e$ and collectively $\nu_x$ as representative of all heavy-lepton neutrinos, comparing the RDF-1.2 (blue lines), RDF-1.8 (red lines), and RDF-1.9 (green lines) hybrid EOS. The quantities are sampled in the comoving frame of reference at a radius of 500~km.}
\label{fig:lumin}
\end{figure*}

\section{Neutrino signal}
\label{sec:neutrinos}
Unlike the $\nu_e$ deleptonization burst, released shortly after the stellar core bounce, the second neutrino burst, associated with the QCD phase transition and the formation of the second shock as a consequence of the PNS collapse, is present in all neutrino flavors. 
In particular, it is dominated by $\bar\nu_e$, which provides ideal prospects for their detection \citep[c.f.][]{Dasgupta10,Fischer18,Balantekin22}. 
Here, we want to focus on these second neutrino bursts and derive connections between the intrinsic properties of these bursts and both the explosion dynamics as well as the EOS. 

Figure~\ref{fig:lumin} shows the evolution of the neutrino luminosities (top panels) and average energies (bottom panels) for all neutrino flavors, 
 sampled at a radius of 500~km in the comoving frame of reference. 
In appendix~\ref{sec:appendix_a} we present an analysis at the level of the optical depth, demonstrating that this radius corresponds indeed to the freely streaming conditions at all simulation times relevant for the release of the second neutrino burst.
Here, $\nu_x$ denotes collectively $\nu_{\mu/\tau}$ while $\bar\nu_{\mu/\tau}$ are omitted for simplicity, for {\tt s30a28} (left panel) and {\tt s40a28} (right panel), as representative cases for all exploding models. 
The times are gauged to the post-bounce time for the release of the second neutrino burst, denoted as $t_{\rm burst}$, for which the values are listed in Table~\ref{tab:fit} for all exploding models under investigation. 
These include also those of previous works, such as simulations with the RDF-1.1 and RDF-1.2 EOS launched from the 50~M$_\odot$ of \citet{Umeda08}, published in \citet{Fischer18} and \citet{Fischer20}, as well as simulations published in \citet{Fischer:2021} launched from the solar metallicity progenitor with a ZAMS mass of 75~M$_\odot$, from the stellar evolution series of \citet{Woosley:2002zz}.

The evolution of the neutrino fluxes and average energies show the behavior that was reported previously from simulations of SN explosions that feature a first-order QCD phase transition, i.e. the canonical post-bounce mass accretion prior to the phase transition with high luminosities on the order of few times $10^{52}$~erg~s$^{-1}$ for all flavors and the sudden rise of the luminosities of all flavors due to the passage of the second shock across the neutrinospheres shortly after the PNS collapse as a direct consequence of the phase transition. 
The second burst is present in all flavors, however, dominated by $\bar\nu_e$ and heavy-lepton flavors, due to the high neutron degeneracy of matter and the low value of $Y_e$. 
Furthermore, while the magnitudes of the $\nu_e$ deleptonization bursts are nearly the same for all SN models, they are determined by the same hadronic EOS and the progenitor structure, the second bursts differ largely in both magnitude as well as width, as can be identified in the inlays in Fig.~\ref{fig:lumin}. 
Models with an early(late) onset of the phase transition, i.e. low(high) onset densities for the phase transition, show a high(low) peak luminosity in the second burst. 
This aspect is the foundation for the following discussion, that gives rise to a number of linear relations.

\begin{figure*}[t!]
\subfigure[]{\includegraphics[angle=0.,width=0.995\columnwidth]{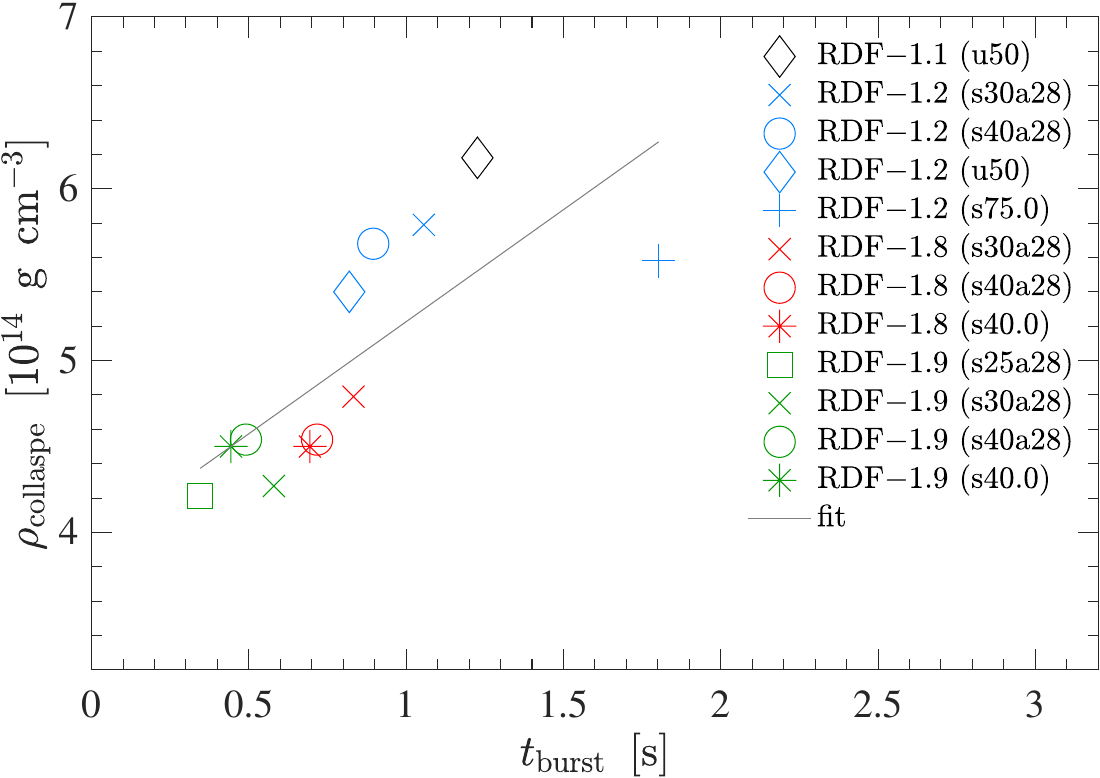}\label{fig:fit_1}}
\hspace{3mm}
\subfigure[]{\includegraphics[angle=0.,width=0.995\columnwidth]{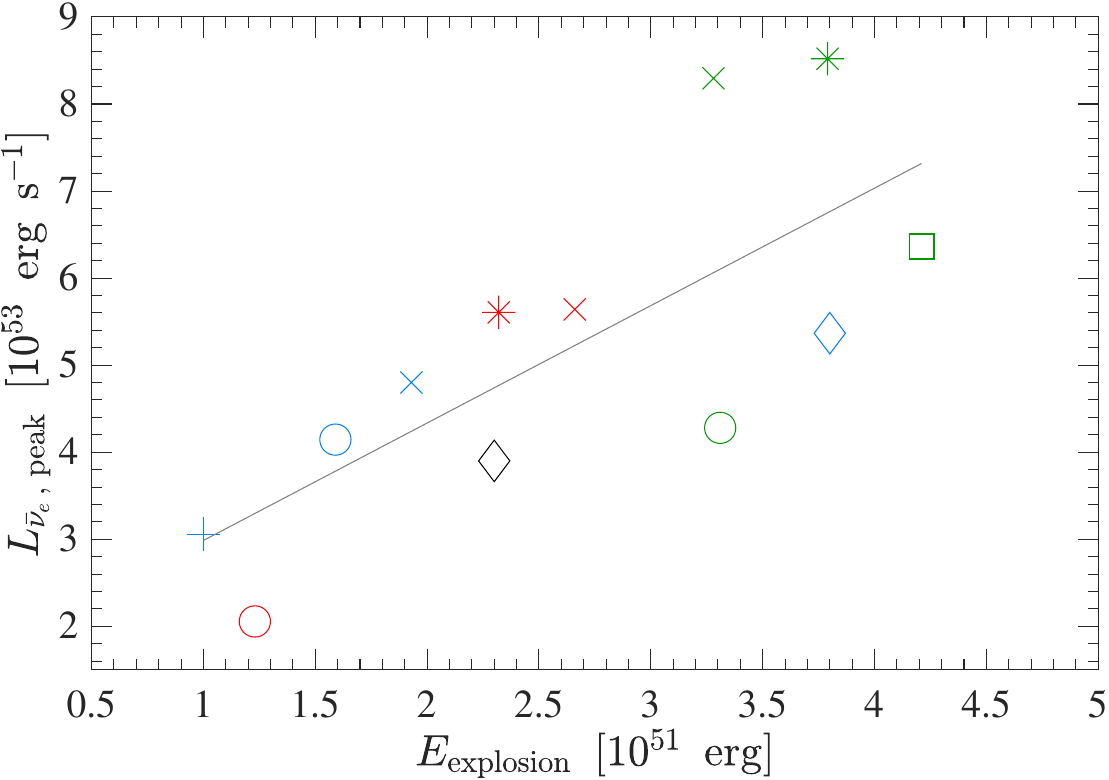}\label{fig:fit_2}}
\\
\subfigure[]{\includegraphics[angle=0.,width=0.995\columnwidth]{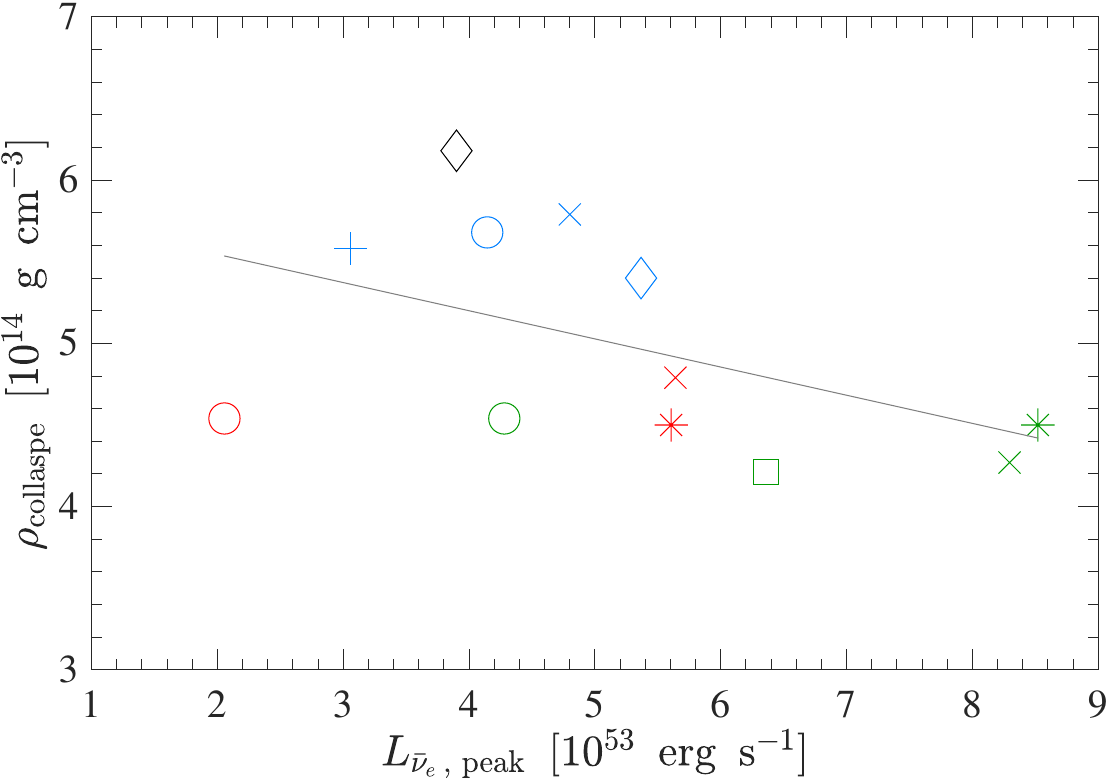}\label{fig:fit_3}}
\hspace{3mm}
\subfigure[]{\includegraphics[angle=0.,width=0.995\columnwidth]{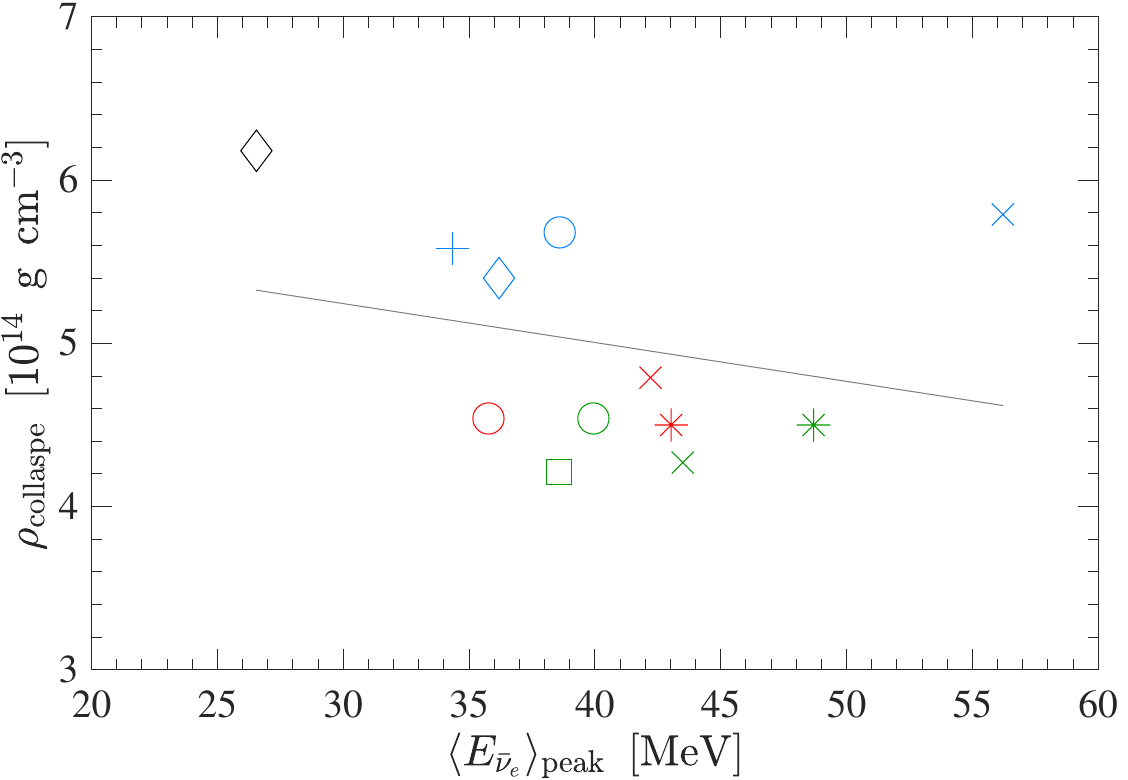}\label{fig:fit_4}}
\\
\subfigure[]{\includegraphics[angle=0.,width=1.0\columnwidth]{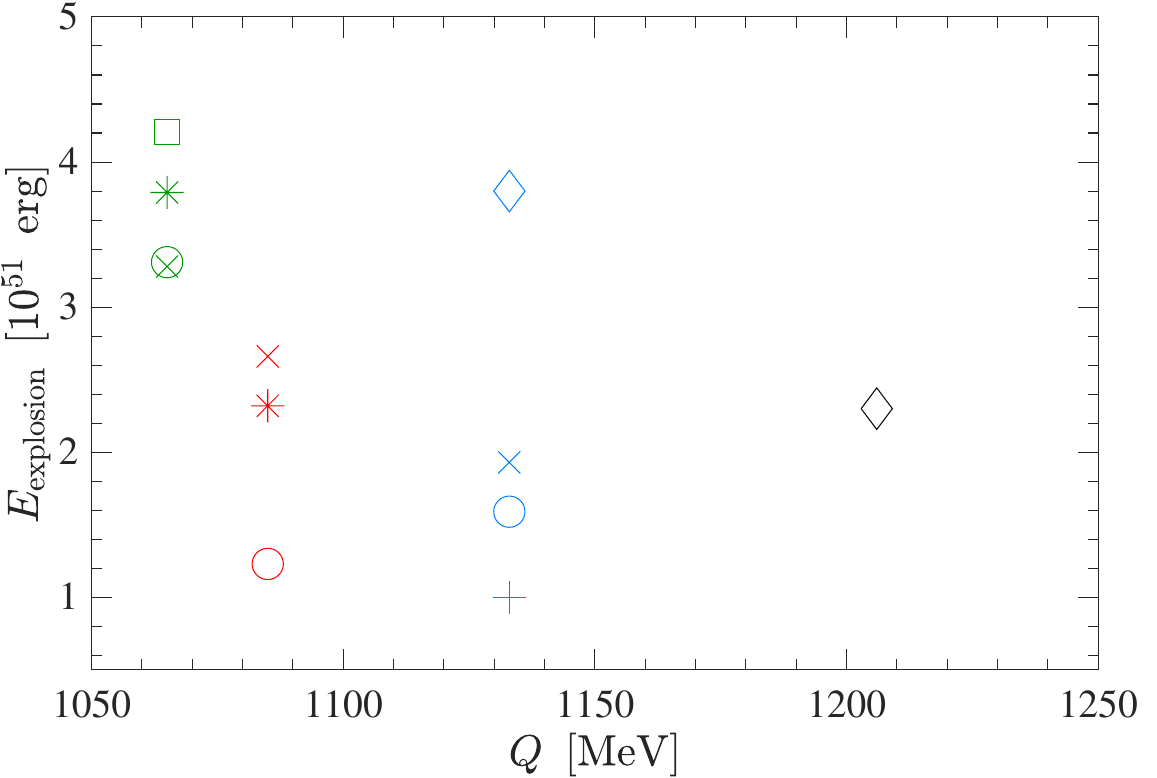}\label{fig:fit_5}}
\hspace{3mm}
\subfigure[]{\includegraphics[angle=0.,width=0.975\columnwidth]{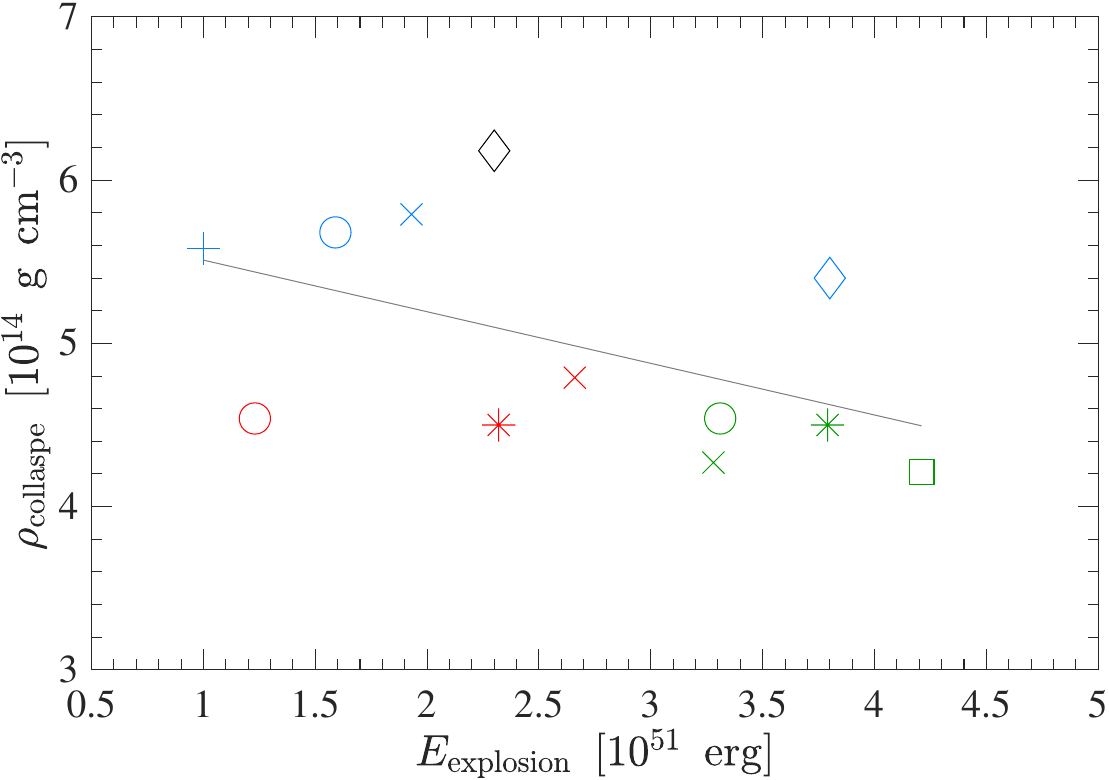}\label{fig:fit_6}}
\\
\\
\caption{Correlations of various quantities and their linear Equations~\eqref{eq:fit1}--\eqref{eq:fit4}, for all exploding models under investigation (see the text for further details). {\em Top panel}:~Onset density for the PNS collapse $\rho_{\rm collapse}$ versus the post-bounce time of the release of the second neutrino burst $t_{\rm burst}$ (left panel) and peak of the $\bar\nu_e$ luminosity $L_{\bar\nu_e,\rm peak}$ versus $E_{\rm explosion}$ (right panel). {\em Middle panel}:~$\rho_{\rm collapse}$ versus $L_{\bar\nu_e,\rm peak}$ (left panel) and $\rho_{\rm collapse}$ versus the peak of the average $\bar\nu_e$ average energy $\langle E_{\bar\nu_e} \rangle_{{\rm peak}}$ (right panel). {\em Bottom panel}:~Comparison of the explosion energy estimates for each explosion model, $E_{\rm explosion},$ and the latent heat, $Q$, of the corresponding $s=3~k_{\rm B}$ EOS (left panel), listed in Table~\ref{tab:EOS}, and the separation of early and late QCD phase transition onset models, corresponding to below and above the $\rho_{\rm collapse}$--$E_{\rm explosion}$ solid line (right panel).}
\label{fig:fit}
\end{figure*}

Further analysis of the large sample of exploding models, including results from previous publications \citep[see][]{Fischer18,Fischer20,Fischer:2021}, reveals several linear relations between observables and EOS quantities, as shown in Fig.~\ref{fig:fit}. These include the relation between the explosion energy $E_{\rm explosion}$ and the peak of the $\bar\nu_e$ luminosity, $L_{\bar\nu_e,{\rm peak}}$, shown in Fig.~\ref{fig:fit_2}.
Further linear relations have been found empirically between the onset density for the PNS collapse $\rho_{\rm collapse}$ and the post bounce time for the release of the second neutrino burst $t_{\rm burst}$, shown in Fig.~\ref{fig:fit_1} as well as between the onset density for the PNS collapse and the peak of the $\bar\nu_e$ luminosity and the $\bar\nu_e$ average energy, shown in Figs.~\ref{fig:fit_3} and \ref{fig:fit_4}, respectively,
\begin{eqnarray}
\rho_{{\rm collapse}} & \simeq & c_1 t_{\rm burst}  + d_1~, \label{eq:fit1}\\
L_{\bar\nu_e,{\rm peak}} & \simeq & c_2 E_{\rm explosion} + d_2~, \label{eq:fit2} \\
\rho_{{\rm collapse}} & \simeq & c_3 L_{\bar\nu_e,{\rm peak}} + d_3~, \label{eq:fit3}\\
\rho_{{\rm collapse}} & \simeq & c_4 \langle E_{\bar\nu_e \rangle,{\rm peak}} + d_4~. \label{eq:fit4}
\end{eqnarray}
The coefficients $c_i$ and $d_i$ are listed in Table~\ref{tab:fit_parameters}. 

A direct connection between the release of the second neutrino burst and the hybrid EOS is the correlation between the post-bounce time for the emission and the onset density. 
EOSs with a high onset density, such as RDF-1.1 and RDF-1.2, feature a late PNS collapse and hence a late second neutrino burst release, whereas the opposite holds for EOS with a low onset density, such as RDF-1.9. The RDF-1.8 EOS is somewhat in between RDF-1.2 and RDF-1.9, as shown in Fig.~\ref{fig:fit_1}. 
The actual values for onset density $\rho_{\rm onset}$ and the PNS collapse $\rho_{\rm collapse}$ are listed in Table~\ref{tab:results} in sec~\ref{sec:systematics}, and the post-bounce times for the release of the burst $t_{\rm burst}$ are given in Table~\ref{tab:fit}. 
This linear dependence between the central density growth and the timescale to reach conditions of the QCD phase transition as well as the PNS collapse, respectively, Equation~\eqref{eq:fit1}, is a consequence of the growth of the PNS mass due to the continuous mass accretion onto the PNS, resulting in the continuous steepening of the gravitational potential, and the incompressibility of matter at high density, given by the nuclear EOS. It is given by the compression modulus and its density dependence, which for SN matter has two contributions, a symmetric matter and asymmetry components \citep[for details, c.f.,][and references therein]{Haensel1979PhLB81}. The former can be expressed as a power law in terms of the baryon density, depending on the nuclear model, while the latter depends on the charge density, i.e. the baryon density and $Y_e$. In RMF models the baryon chemical potential can be expressed in terms of the quasi-particle Fermi gas contributions as well as interaction contributions due to the mesons, from where the dependence on density can be understood semi-analytically, which has been shown in \citet{Dexheimer2008PhRvC77}.
The outlier here is {\tt s75.0}, which has a much lower mass accretion rate than any other model explored in this study. This results in a spread of the onset density in the range of $\rho_{\rm onset}\simeq 5.4 - 5.8 \times 10^{14}$~g~cm$^{-3}$ for the RDF-1.2 models, and an overall uncertainty of $\rho_{\rm onset}\simeq 4.2 - 6.2 \times 10^{14}$~g~cm$^{-3}$ across all simulations, with an error well within 95\% confidence level from the linear fit for any model. 
\begin{table*}[ht!]
\begin{center}
\caption{ Summary of the Explosion Models.}
\begin{tabularx}{0.99\textwidth}{l@{\hspace{1cm}}c@{\hspace{1cm}}c@{\hspace{1cm}}c@{\hspace{1cm}}c@{\hspace{1cm}}c@{\hspace{1cm}}c@{\hspace{1cm}}cc}
\hline
\hline
Progenitor & EOS & $t_{\rm burst}^a$ & $L_{\bar\nu_e, {\rm peak}}^b$ & $\langle E_{\bar\nu_e} \rangle_{\rm peak}^c$ & $E_{\rm expl}^d$ & $E_{\bar\nu_e}^{{\rm burst},e}$ & $N_{\bar\nu_e}^{{\rm burst}, f}$ \\
& RDF & $[$s$]$ & $[10^{53}\,\,{\rm erg\,\,\,s}^{-1}]$ &$[$\rm{MeV}$]$ &  $[10^{51}\,\,{\rm erg}]$ & $[10^{51}\,\,{\rm erg}]$ & $[10^{55}]$  \\
\hline
{\tt s25a28} & 1.9 & 0.345 & 6.36 & 38.59 &  4.21 &  1.25  &  4.20  \\
{\tt s30a28} & 1.2 & 1.056 & 4.80 & 56.21 &  1.93 &  1.41  &  5.08  \\
{\tt s30a28} & 1.8 & 0.833 & 5.64 & 42.21 &  2.66 & 1.12   &  3.91 \\
{\tt s30a28} & 1.9 & 0.580 & 8.30 & 43.49 &  3.28 &  1.72  &  6.14 \\
{\tt s40a28} & 1.2 & 0.895 & 4.15 & 38.60  &  1.59 &   1.53 &  5.45 \\
{\tt s40a28} & 1.8 & 0.717 & 2.06 & 35.77 & 1.23 &  1.53  &  5.15 \\ 
{\tt s40a28} & 1.9 & 0.491 &  4.28 & 39.94 & 3.31 &  1.79 &  6.46 \\
{\tt s40.0}  & 1.8 & 0.694 & 5.61 & 43.03 &  2.32 &  1.97  &  6.07 \\
{\tt s40.0}  & 1.9 & 0.443 &  8.52 & 48.69 &  3.79 &  1.95  &  7.17 \\
\hline
{\tt u50}$^g$ & 1.1 & 1.227 & 3.90 & 26.55 & 2.3 &  1.85 &  6.8 \\
{\tt u50}$^h$ & 1.2 & 0.819 & 5.37 & 36.19 & 3.8 &  2.44 &  9.3 \\
\hline
{\tt s75.0}$^i$ & 1.2 & 1.803 & 3.06 & 34.35 & 1.0 &  1.06 &  3.9 \\
\hline
\label{tab:fit}
\end{tabularx}
\end{center}
{\bf Notes.}\\
$^a$~Post-bounce time of the second neutrino burst release, sampled in the comoving frame at 500~km, when $L_{\bar\nu_e}>10^{53}$~erg~s$^{-1}$. \\
$^b$~Maximum value of the $\bar\nu_e$ luminosity in the burst.  \\
$^c$~ Maximum value of the average $\bar\nu_e$ energy in the burst. \\
$^d$~Diagnostic explosion energy. \\
$^e$~ Integrated energy of $\bar\nu_e$ in the second burst, starting between the $L_{\bar\nu_e}$ rise and when $L_{\bar\nu_e}\leq 10^{52}$~erg~s$^{-1}$. \\
$^f$~ Integrated number of $\bar\nu_e$ in the second burst, same as $E_{\bar\nu_e}^{\rm burst}$. \\
$^g$~Data from \citet{Fischer18}, launched from 50~M$_\odot$ progenitor \citep{Umeda08}. \\
$^h$~Data from \citet{Fischer20}, launched from 50~M$_\odot$ progenitor \citep{Umeda08}. \\
$^i$~Data from \citet{Fischer:2021}, launched from 75~M$_\odot$ progenitor \citep{Woosley:2002zz}. 
\end{table*}
\\
\\
\begin{table}[htp]
\caption {Linear fitting parameters for Eqs.~\eqref{eq:fit2}--\eqref{eq:fit4}.}
\begin{tabularx}{0.47\textwidth}{l@{\hspace{1cm}}c@{\hspace{1cm}}c@{\hspace{1cm}}c}
\hline
\hline
Dependencies & Label & $c$ & $d$ \\
\hline
$\rho_{{\rm collapse}}(t_{\rm burst})$ & 1 & 1.304 & 3.922 \\
$L_{\bar\nu_e,{\rm peak}}(E_{\rm explosion})$ & 2 & 1.349 & 1.638 \\
$\rho_{{\rm collapse}}(L_{\bar\nu_e,{\rm peak}})$ & 3 & -0.172 & 5.890 \\
$\rho_{{\rm collapse}}(\langle E_{\bar\nu_e \rangle ,{\rm peak}})$ & 4 & -0.238 & 5.959 \\
\\
\hline
\end{tabularx}
\label{tab:fit_parameters}
\end{table}

The evolution of the central fluid element's density prior to the phase transition is determined by the hadronic EOS, for which we explore the DD2 class of RMF EOS here. 
Variations employed here between the original DD2 and its supersaturation density softer variation DD2F, result in somewhat higher central densities during the SN post-bounce evolution, which concerns RDF-1.1 and RDF-1.2 in Fig.~\ref{fig:fit_1}, in comparison to the DD2 based models RDF-1.8 and RDF-1.9, featuring systematically lower onset densities and hence shorter timescales for the onset of quark matter.

In order to determine this correlation quantitatively, one has to take the progenitor model dependence into account,   which determines the mass accretion rate and hence the timescale for the PNS mass to grow. 
Comparing {\tt s30a28}, {\tt s40a28}, and {\tt s40.0}, even though this correlation holds qualitatively, the appearance of the second bursts varies by several hundreds of milliseconds, despite similar central PNS densities for the hadron-quark phase transition as well as the PNS collapse for RDF-1.2, RDF-1.8, and RDF-1.9 (see Table~\ref{tab:results} in sec~\ref{sec:systematics}). 
The reason is the different post-bounce evolution prior to the phase transition. 
Relevant here is the higher temperature obtained for {\tt s40a28}, in combination with the strong temperature dependence of the onset densities for the phase transition within the RDF EOS \citep[c.f.][]{Bastian:2021}, i.e. the onset density shifts toward lower values for increasing temperatures. 
Consequently, the second neutrino burst is launched earlier for models that feature a higher temperature during the post-bounce evolution prior to the onset of the phase transition. 

Comparing the $\bar\nu_e$ luminosity peaks of the second neutrino burst with the diagnostic explosion energies, both listed in Table~\ref{tab:fit} and plotted in Fig.~\ref{fig:fit_2}, a linear correlation is assumed too.
This is related to the energetics of the second shock. 
 Models with a high(low) explosion energy correspond to the developments of stronger(weaker) second shocks, in terms of high(low) shock expansion velocities as well as high(low) matter velocities in the post-shock layer. Consequently, matter ahead of the expanding shock experiences stronger(weaker) shock heating. This, in turn, results in higher(lower) matter temperatures in the post-shock layers for models with high(low) explosion energies, and hence, higher(lower) neutrino luminosities in the second burst. 
The assumed empirical relation between explosion energy estimate, $E_{\rm explosion}$, and the peak $\bar\nu_e$ luminosity expression, Equation~\eqref{eq:fit2}, is shown as the gray solid line in Fig.~\ref{fig:fit_2}.
The neutrino luminosities are sampled in the comoving frame of reference at 500~km.
In Appendix~\ref{sec:appendix_a} we perform an analysis of the neutrino decoupling, to confirm that this sampling radius corresponds to the free-streaming regime. 
Note the outliers in Fig.~\ref{fig:fit_2}, RDF-1.8 {\tt s40s28} (red open circle) and RDF-1.9 {\tt s30a28} and {\tt s40.0} (green $*$ and ${\rm x}$ symbols).
These have the largest deviation from the linear fit~\eqref{eq:fit2}, however, the deviation from the fit is well within the 95\% confidence level. 

 Combining the two linear correlations of Equations~\eqref{eq:fit1} and \eqref{eq:fit2} leads to the relation between the explosion energy estimate and the onset density of quark matter, $E_{\rm explosion}$ and $\rho_{\rm onset}$, respectively.
This relation is shown in Fig.~\ref{fig:fit_6}.
The analogy has already been reported in the original work of \citet{Fischer18} that the more delayed the phase transition occurs, corresponding to high onset densities for the appearance of quark matter, the less energetic the subsequent explosions are.
Connecting this relation with the relation between explosion energy estimate and the peak of the $\bar\nu_e$ luminosity empirically, results in the relation between the onset density of the PNS collapse, $\rho_{\rm onset}$, and $L_{\bar\nu_e\,,\,{\rm peak}}$, shown in Fig.~\ref{fig:fit_3}.
The linear fit assumed, Equation~\eqref{eq:fit3}, is shown via the solid gray line.
The model with the largest deviation is {\tt s40a28} RDF-1.8 (open red circle), however, well within 1 standard deviation.

 Furthermore, since the peak of the $\bar\nu_e$ luminosity cannot be directly related to possible observations, a model for the reconstruction of the luminosity would be needed, in addition, we examine the average neutrino energy.
Precisely, the peak of the average $\bar\nu_e$ energy of the second neutrino burst, denoted as $\langle E_{\bar\nu_e} \rangle_{\rm peak}$. 
The values are listed in Table~\ref{tab:fit} for all explosion models and the data are shown in Fig.~\ref{fig:fit_4}, for which the linear relation~\eqref{eq:fit4} between $\langle E_{\bar\nu_e} \rangle$ and $\rho_{\rm collapse}$ has been assumed empirically. 
In Table~\ref{tab:results}, we also list the values of the total energy and number of $\bar\nu_e$ emitted in the burst, $E_{\bar\nu_e}^{\rm burst}$ and $N_{\bar\nu_e}^{\rm burst}$, respectively, which might be useful for future analysis regarding neutrino detection prospects. The quantities are integrated over the time interval of the release of the second neutrino burst, starting when the $\bar\nu_e$ luminosity, sampled in the comoving frame at 500~km, starts to rise sharply, until it drops below a value of around $10^{52}$~erg~s$^{-1}$.

For the deeper analysis of the physical origin of the linear relations, Equations~\eqref{eq:fit1}--\eqref{eq:fit4}, between neutrino observables and EOS bulk properties, we compare the $s=3~k_{\rm B}$ latent heat $Q$ for all EOSs of the explosion models, as listen in Table~\ref{tab:EOS} (see also Table~\ref{tab:fit}), with the explosion energy estimates. 
 The results are shown in Fig.~\ref{fig:fit_5}, from which it becomes evident that EOS with small latent heat, such as RDF-1.1 (black marker) and RDF-1.8 (red markers), have generally low explosion energies.
The latent heat is expected to be a valid measure of the explosion dynamics as it represents the energy that is potentially released during the phase transition.
Even though naively one might think the opposite appears to be true too, i.e. EOS with large latent heat feature high explosion energies, such as RDF-1.9 (green markers), the data cannot generally confirm this tendency.
There are clear outliers beyond a $1 \sigma$ margin, RDF-1.2 (blue markers) for the runs {\tt s75.0}, {\tt s30a28}, and {\tt s40a28}. 
The nonlinearity of the radiation hydrodynamical phenomena associated with the phase transition, such as the supersonic PNS collapse, formation of a shock wave, associated energy transport as well as neutrino losses, make the situation mode complex such that a simple one-to-one correspondence between the EOS latent heat and the explosion energy estimate cannot be confirmed generally.
Note also that different definitions of the latent heat \citep[c.f.][]{Lope-Oter2022PhRvC.105e2801L} result not only in qualitative different values for $Q$ but also in a different relation between $Q$ and the explosion energy estimates. 
This might be convoluted even further due to the varying onset densities for the phase transition, with low onset densities for RDF-1.8 and RDF-1.9 and higher onset densities for RDF-1.1 and RDF-1.2, and the linear relation between the explosion energy estimate and the onset density.

Models with an early explosion onset lie below the solid line shown in Fig.~\ref{fig:fit_6}, which corresponds to the linear fit of $\rho_{\rm collapse}$ and $E_{\rm explosion}$ for all explosion models under investigation.

 However, further investigation of the latent heat does not reveal any correlations between neutrino observables, e.g., $E_{\bar\nu_e}^{\rm burst}$ and $N_{\bar\nu_e}^{\rm burst}$. 
In Appendix~\ref{sec:appendix_b} we have therefore performed a neutrino pulse propagation analysis, similar to the one performed by \citet{Liebendorfer04}, sampling the neutrino pulse at different radii, in order to confirm the integrated quantities remain unchanged despite the numerical diffusion of the luminosity peak due to mesh effects.


Note that equivalently to the onset density for the PNS collapse, one can choose the onset density for the phase transition as both densities are related (see Table~\ref{tab:results}). Values for the onset densities for the models {\tt u50} are taken from \citet{Fischer18} for RDF-1-1, $\rho_{\rm collapse}=6.2\times 10^{14}$~g~cm$^{-3}$, from \citet{Fischer20} for RDF-1.2, $\rho_{\rm collapse}=5.4\times 10^{14}$~g~cm$^{-3}$, and for {\tt s75.0} from \citet{Fischer:2021} using RDF-1.2, $\rho_{\rm collapse}=5.6\times 10^{14}$~g~cm$^{-3}$. 


\section{Summary and Conclusions}
\label{sec:summary}
Microscopic hadron-quark hybrid matter EOS of \citet{Bastian:2021}, featuring a first-order hadron-quark phase transition, from the DD2 RMF EOS class as well as the softer DD2F and DD2Fev variations, are employed in general relativistic neutrino radiation hydrodynamic simulations of CCSN, exploring the dependence on the underlying EOS and the stellar models. 
 Note that while the results reported in the present paper are unlikely to alter qualitatively, they are likely to vary quantitatively when constructing a phase transition from a different hadronic model.
However, since the hadronic EOS is presently still highly uncertain, in particular at densities in excess of nuclear saturation density and at large isospin asymmetry, model EOSs are selected that are compatible with astrophysical constraints, with all of which presently the DD2 EOS class agrees quantitatively very well. 
We leave the exploration of the dependence on the hadronic EOS for a future study.

Progenitors are selected from two different stellar evolution calculations, with ZAMS masses of 25--40~M$_\odot$, with differences in the core structures, in particular the density in the silicon-sulfur layers, resulting in different post-bounce evolutions. 
This has a direct impact on the appearance of the QCD phase transition. 

Models with low(high) post-bounce mass accretion rate result in the slow(fast) growth of the enclosed PNS mass. 
This aspect is critical. 
It results in conditions in which the enclosed PNS mass exceeds the maximum mass, which is given by the hybrid EOS, for most EOS under investigation at the moment when the PNS becomes gravitationally unstable due to entering the hadron-quark coexistence region. 
The latter is characterized by a substantially reduced adiabatic index. 
Consequently, the PNS collapse results in black hole formation, for which two scenarios are found. 
One, prompt collapse in which the black hole forms before the second shock breaks out, and two, with the expansion of the second shock before the central PNS collapses into a black hole. 
Note that in general it is confirmed that there is a delay from the onset of the phase transition, identified in the simulations when finite values of the quark matter volume fraction are obtained, and the subsequent PNS collapse, up to several 100~ms. 

In cases when the enclosed PNS mass remains below the maximum mass of the hybrid EOS, stable PNS remnants are obtained after the PNS collapse, now featuring a quark matter core. 
These are initially very massive, exceeding 1.5~M$_\odot$ for all EOSs under investigation due to the high temperatures and the strong temperature dependence of the phase boundaries for the RDF class of hybrid EOS. 
Later, during the PNS deleptonization phase, when the temperature decreases due to the emission of neutrinos of all flavors, the quark core masses approach the cold $\beta$-equilibrium limits.

It remains to be shown whether positive explosion energies might be obtained for the failed models that belong to the delayed scenario, for which it will be required to simulate beyond the appearance of the event horizon, more precisely the apparent horizon~\citep[][]{Rahman2022MNRAS512}. 
Such models might be candidates for the collapsar scenario, i.e. the presence of a black hole in the center, while the explosion proceeds as the second shock wave continues to expand to increasingly larger radii. 
This idea has long been investigated in the context of rotational magnetized failed SN \citep[c.f.][and references therein]{MacFadyen99,Proga03,Ott11,Aloy20}, often in connection with the emission of gamma-ray bursts. 

For all explosion models, a second neutrino burst is released, which is absent in any other SN {\rm explosion} scenario. 
The exception might be the scalarization of bosonic degrees of freedom \citep[c.f.][]{Kuroda2023}. 
In cases of the first-order QCD phase transition, the second neutrino burst emission is associated with the propagation of the second shock across the neutrinospheres. 
The present paper establishes a number of phenomenological linear relations between signatures of the second neutrino burst, in particular for $\bar\nu_e$, and the explosion dynamics as well as the EOS. 
It enables the determination of the onset density for quark matter from a future observation of the next galactic SN neutrino signal, from the peak of the $\bar\nu_e$ luminosity and the average energies. 
On the contrary, the absence of a second neutrino burst on timescales below or around one or a few seconds after the onset of neutrino detection will provide a lower bound for the onset density of quark matter, i.e. providing a constraint for the QCD phase diagram for SN matter featuring $Y_e\simeq 0.2-0.3$, with $\rho_{\rm onset}\gtrsim 4\times 10^{14}$~g~cm$^{-3}$ and in the temperature range of $T\simeq40\pm10$~MeV. 
The present analysis covers a wide range of possible degeneracies.
It includes the post-bounce mass accretion phase, due to different stellar models and the hadronic EOS in terms of the DD2 RMF EOS, including the variations DD2F and DD2Fev, both of which represent softer supersaturation density EOS than DD2. 
It is worth noting that it becomes increasingly difficult conceptually to construct first-order phase transitions to the RDF class of quark matter models from soft hadronic EOS and fulfill all present nuclear physics and astrophysical constraints simultaneously.
Furthermore, several aspects of the quark matter EOS, such as onset densities, the magnitude of the density jump as a consequence of the phase transition construction, improved quark matter descriptions, e.g., that include pairing and the presence of color-superconducting phases \citep[c.f.][]{Blaschke05,Ruester05,Ivanytskyi22EPJA,Ivanytskyi22Part} as well as other phase transition constructions \citep[][]{Maslov19}, are yet incompletely understood. 
These aspects, however, extend beyond the scope of the present study and are left for future explorations.
Note that here $\bar\nu_e$ are selected as observable since they have the highest prospects for detection via the inverse beta decay through the current generation of operating water Cherenkov detectors. 

The present analysis demonstrates that multi-messenger SN observables are an alternative route to constrain the dense matter EOS, with the presence and the intrinsic properties of a nonstandard observable second neutrino burst, around or less than 1s after the onset of neutrino detection. It is complementary to future planned heavy-ion collider programs such as FAIR Facility for Antiproton and Ion Research at GSI in Darmstadt (Germany) and NICA (Nuclotron Ion Collider) in Dubna (Russia), which aim to probe the QCD phase diagram in the baryon rich regime, however, at somewhat lower isospin asymmetry.

\section*{acknowledgments}
We would like to thank Kei~Kotake, Shunsaku~Horiuchi, Shota~Shibagaki, and Takami~Kuroda for inspiring discussions and helpful comments.
The authors acknowledge support from the Polish National Science Center (NCN) under grant No. 2020/37/B/ST9/00691 (T.F., N.K.L.). All computer simulations were performed at the Wroclaw Center for Scientific Computing and Networking (wcss.pl).

\section*{ORCID iDs}
\noindent Noshad Khosravi Largani https://orcid.org/0000-0003-1551-0508 \\
Tobias Fischer https://orcid.org/0000-0003-2479-344X \\
Niels Uwe F. Bastian http://orcid.org/0000-0001-9793-240X
\clearpage
\appendix

\section{Radiation Pulse Propagation}
\label{sec:appendix_b}
\begin{figure}[htp]
\centering
\includegraphics[angle=0.,width=1.0\columnwidth]{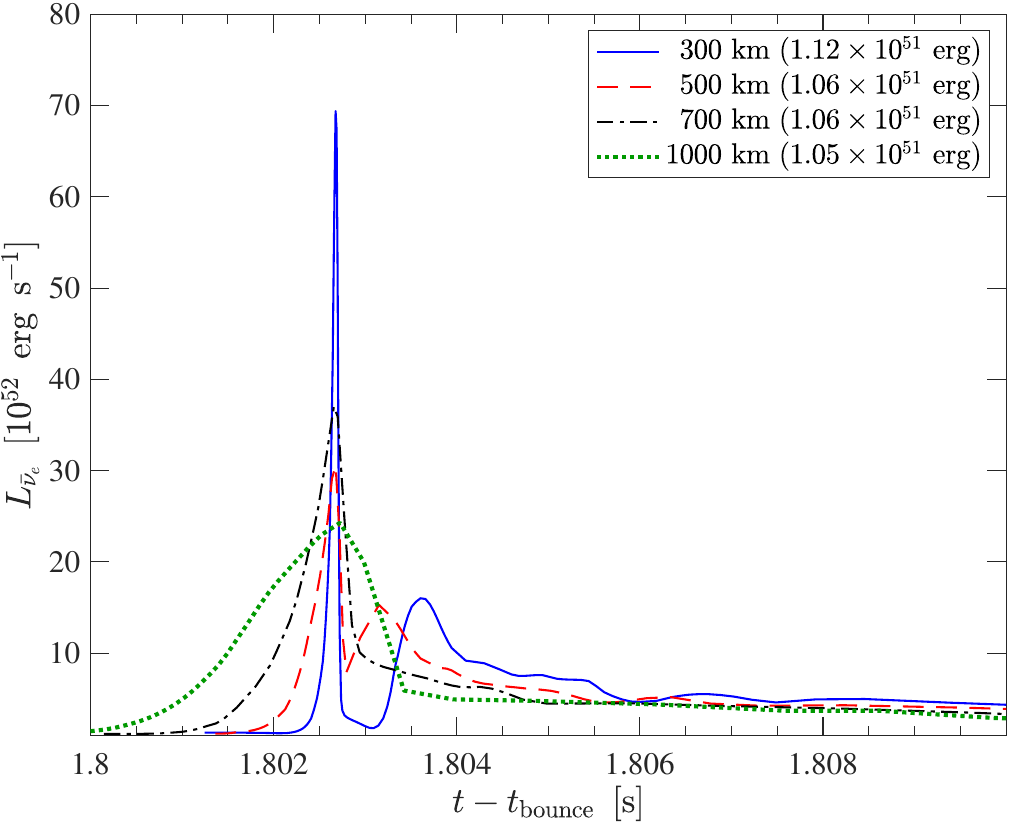}
\caption{Release of the $\bar\nu_e$ burst sampled at four selected radii of $R=300$~km (solid blue line), 500~km (dashed red line), 700~km (back dashed-dotted line) and 1000~km (green dotted line), for the model of \citet{Fischer:2021} launched from the 75~M$_\odot$ progenitor of solar metalicity from the stellar evolution series of \citet{Woosley:2002zz}.  Note that all post-bounce times are shifted by the corresponding sampling radii, i.e. $R/c$, such that the peaks of the neutrino bursts center at the post-bounce time. The legend contains the data for the neutrino energy release in the burst (see the text for details).}
\label{fig:neutrino-burst}
\end{figure}

To ensure the robustness of the results presented here, we perform an analysis on the dependence of the sampling radius for the neutrino observables. 
Therefore, we revisit an old problem related to the diffusion of radiation bursts as they propagate across the mesh \citep[c.f. sec.~4.4 in][and references therein]{Liebendorfer04}. 
The sampling radii are varied in the comoving frame for the determination of the neutrino fluxes and spectra, between  $R=300, 500, 700,$~and $1000$~km, for the explosion model of \citet{Fischer:2021}, which was launched from the solar metalicity 75~M$_\odot$ progenitor of \citet{Woosley:2002zz}. 
The results are shown in Fig.~\ref{fig:neutrino-burst}, 
 as a function of post-bounce shifted with respect to $R/c$ such that the peaks coincide,
together with the integrated $\bar\nu_e$ energy emitted in the burst, for which we integrate the $\bar\nu_e$ luminosities with respect to time, starting at the moment when the $\bar\nu_e$ luminosity has its minimum just before the sudden rise, until the time when the luminosity drops below around $10^{52}$~erg~s$^{-1}$. 
The corresponding values are given in the legend, from which it becomes evident that a sampling radius of 300~km, with $E_{\bar\nu_e,\,{\rm burst}}=1.12\times 10^{51}$~erg, still does not correspond to the fully freely streaming regime.

The values agree quantitatively for larger radii, corresponding $E_{\bar\nu_e,\,{\rm burst}}=1.06\times 10^{51}$~erg. 
Larger radii, 1000~km, face the problem of the numerical diffusion of the burst during its propagation, where $E_{\bar\nu_e,\,{\rm burst}}=1.05\times 10^{51}$~erg. 
We also perform the same analysis for the integration of the $\bar\nu_e$ number fluxes $\dot N_{\bar\nu_e}$ sampled at the same radii, for which we obtain a similar agreement for the larger sampling radii, i.e. $3.9819\times 10^{55}$ at 500~km and $3.8849\times 10^{55}$ at 700~km. 
A similar agreement is expected to hold qualitatively for all other explosion models, featuring the same phenomenology of the second shock propagation across the neutrinospheres regarding neutrino decoupling.

\section{Neutrino decoupling and optical depth}
\label{sec:appendix_a}
To guarantee that the sampling radius for the determination of the neutrino luminosities and average energies are located outside the neutrinospheres at all times of the simulations, we perform an analysis at the level of the energy-dependent neutrinosphere, for one of the explosion models, {\tt s40a28} RDF-1.2, representative for all explosion models under investigation. 
Therefore, we compute the neutrino energy optical depth, $\tau$, by integrating the local neutrino energy-dependent inverse transport mean-free path, weighted by the neutrino distribution function, from the surface of the computational domain of the SN simulations to the center. 
Here, we follow the description of \citet{Fischer12} in Section~II~B. 
The neutrinospheres are then defined as $R_\nu=R(\tau=2/3)$.
Note that the transport mean-free path contains contributions from all weak interaction channels, charged current, elastic neutrino-nucleon/nucleus scattering, inelastic neutrino electron/positron scattering, and pair processes. 

\begin{figure*}[htp]
\centering
\subfigure[~800~ms post bounce, before PNS collapse]{\includegraphics[angle=0.,width=1.975\columnwidth]{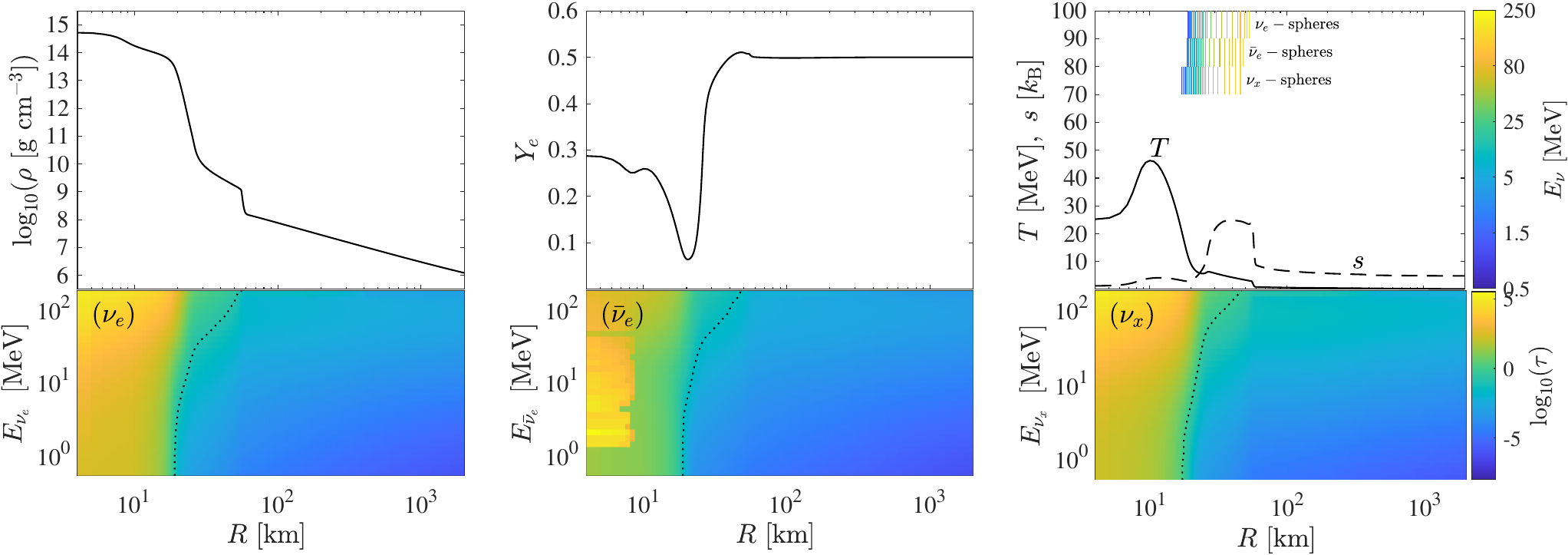}\label{fig:moments_a}}
\subfigure[~896~ms post bounce, during the initial shock propagation after PNS collapse]{\includegraphics[angle=0.,width=1.975\columnwidth]{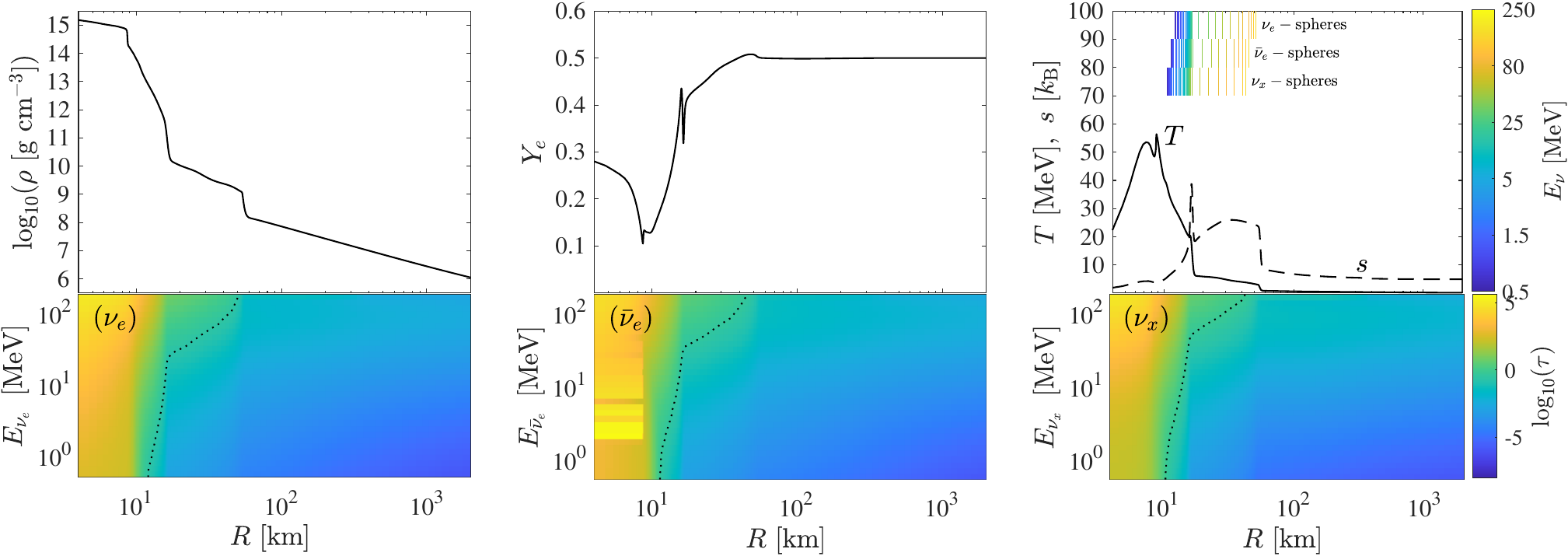}\label{fig:moments_b}}
\subfigure[~899~ms post bounce, during shock propagation across the neutrinospheres]{\includegraphics[angle=0.,width=1.975\columnwidth]{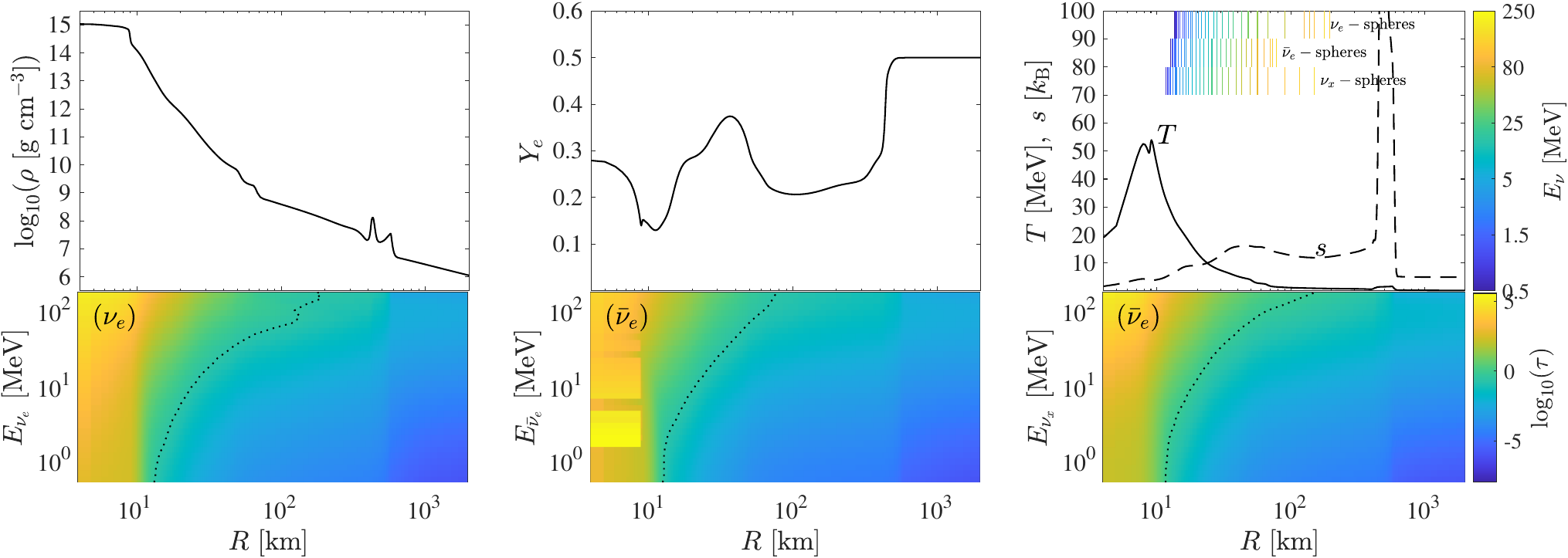}\label{fig:moments_c}}
\caption{Top panels:~Radial profiles of selected hydrodynamic quantities at three selected post-bounce times, showing (from left to right) the rest-mass density, $Y_e$ as well as temperature (solid lines) and entropy per baryon (dashed lines). The color-coded vertical lines in the upper right panels mark the locations of the energy-dependent neutrinospheres for all three flavors, where the color coding corresponds to the neutrino energy. Bottom panels:~Radial profiles of the neutrino energy-dependent logarithm of the optical depth, $log_{10}(\tau)$, for (from left to right) $\nu_e$, $\bar\nu_e$ and $\nu_x\equiv\nu_{\mu/\tau}$ (collectively for all heavy-lepton flavors. The black dashed lines mark the condition, $\tau=2/3$, i.e. the location of the corresponding neutrinospheres.}
\label{fig:moments}
\end{figure*}

The results of this analysis are shown in Fig.~\ref{fig:moments}, where we display radial profiles of selected quantities (top panels), rest-mass density, $Y_e$, temperature and entropy (from left to right), and the energy-dependent optical depths in log-scale (bottom panels), separated into (from left to right) $\nu_e$, $\bar\nu_e$ and $\nu_x\equiv\nu_\mu$, collectively for all heavy-lepton flavors, sampled at three different post-bounce times.
Fig.~\ref{fig:moments_a} corresponds to typical post-bounce situation, despite the central fluid elements having entered the hadron-quark mixed phase, the neutrinospheres follow the canonical hierarchy, $R_{\nu_e}<R_{\bar\nu_e}\leq R_{\nu_x}$, which are illustrated by the colored vertical lines in the upper right panels of Fig.~\ref{fig:moments} with the color coding corresponding to the neutrino energy.
To support these findings, the bottom panels, containing $\tau$, quantify the transition from optically thin (transparent) where $\tau$ is small to the opaque regime where $\tau$ is large. 
To guide the eye, the black dashed lines mark the location where $\tau=2/3$, which is the canonical value for the definition of the corresponding neutrinosphere radii. 

From this analysis it is evident that at a radius of 500~km, where we sample the neutrino observables, correspond to the freely streaming conditions, also at later times shown in Figs.~\ref{fig:moments_b} and \ref{fig:moments_c}. 
The latter two are especially crucial since they belong to the phase of the initial and later shock propagation, in particular across the neutrinospheres, i.e. when the second neutrino burst is launched. Previous analyses, that sampled the neutrino fluxes and spectra at radii of 200--300~km are to be revised, since, as shown in Fig.~\ref{fig:moments_c}, they belong to the region where the high-energy tail of the neutrino distribution is still not fully freely streaming.

\bibliographystyle{aasjournal}
\bibliography{references}

\end{document}